\documentclass[11pt,letterpaper]{article}%
\usepackage{makeidx}
\usepackage{eurosym}
\usepackage{amsthm}
\usepackage{graphicx}
\usepackage{amsmath, bm}
\usepackage{amssymb}
\usepackage{amsfonts}
\usepackage{euscript}
\usepackage{natbib}
\usepackage{geometry}
\usepackage{algorithm}
\usepackage[usenames]{color}
\usepackage{xr}
\usepackage{tabularx}
\usepackage{amsmath}%
\providecommand{\U}[1]{\protect \rule{.1in}{.1in}}
\newtheorem{theorem}{Theorem}

\newtheorem{assumption}{Assumption}
\newtheorem{corollary}{Corollary}
\newtheorem{lemma}{Lemma}

\newtheoremstyle{ecta}
{\medskipamount}{\bigskipamount}{\normalfont}{1.4em}{\scshape}{:}{1em}{}
\theoremstyle{ecta}
\newtheorem*{example*}{Example}

\newcolumntype{Y}{>{\centering \arraybackslash}X}

\allowdisplaybreaks

\begin{document}

\title{Standard Errors When a Regressor is Randomly Assigned}
\author{Denis Chetverikov\thanks{Department of Economics, UCLA, Los Angeles, CA
90095-1477 USA. Email:\ chetverikov@econ.ucla.edu}\\UCLA
\and Jinyong Hahn\thanks{Department of Economics, UCLA, Los Angeles, CA 90095-1477
USA. Email:\ hahn@econ.ucla.edu}\\UCLA
\and Zhipeng Liao\thanks{Department of Economics, UCLA, Los Angeles, CA 90095-1477
USA. Email:\ zhipeng.liao@econ.ucla.edu}\\UCLA
\and Andres Santos\thanks{Department of Economics, UCLA, Los Angeles, CA 90095-1477
USA. Email:\ andres@econ.ucla.edu}\\UCLA}
\date{\today }
\maketitle

\begin{abstract}
We examine asymptotic properties of the OLS estimator when the values of the regressor of interest are assigned randomly and independently of other regressors. We find that the OLS variance formula in this case is often simplified, sometimes substantially. In particular, when the regressor of interest is independent not only of other regressors but also of the error term, the
textbook homoskedastic variance formula is valid even if the error term and auxiliary regressors exhibit a general dependence structure. In the context of randomized controlled trials, this conclusion holds in completely randomized
experiments with constant treatment effects. When the error term is heteroscedastic with respect to the regressor of interest, the variance formula has to be adjusted not only for heteroscedasticity but also for correlation structure of the error term. However, even in the latter case, some simplifications are possible as only a part of the correlation structure of the error term should be taken into account. In the context of randomized control trials, this implies that the textbook homoscedastic variance formula is typically not valid if treatment effects are heterogenous but heteroscedasticity-robust variance formulas are valid if treatment effects are independent across units, even if the error term exhibits a general dependence structure. In addition, we extend the results to the case when the regressor of interest is assigned randomly at a group level, such as in randomized control trials with treatment assignment determined at a group (e.g., school/village) level.

\bigskip

\noindent JEL Classification: C14, C31, C32\bigskip

\noindent \textit{Keywords:} Cluster Robust Inference; Randomized Control Trial

\end{abstract}

\section{Introduction\label{sec:intro}}

Textbook discussion of linear regression usually begins with a standard model of the form $\mathbf{Y}=\mathbf{X}\theta+\bm{\epsilon}$,\ where it is assumed that $\mathbf{X}$ is a nonstochastic matrix (with full column rank) of regressors and the error vector $\bm{\epsilon}$ has mean zero and variance matrix proportional to an identity matrix. 
As is well known, such an assumption justifies the formula $s^{2}\left( \mathbf{X}^{\top}\mathbf{X}\right)^{-1}$ as an estimator of the variance of the OLS estimator, where $s^{2}$ is equal to the sum of squares of the estimated residuals divided either by the sample size or the degrees of freedom. 
This formula is easy to use but, as is typically taught, may not be valid if the variance matrix $\Omega$ of the error vector $\bm{\epsilon}$ is not proportional to an identity matrix.
In such cases, the variance of the OLS estimator should be based on the formula $\left(  \mathbf{X}^{\top}\mathbf{X}\right)  ^{-1}\mathbf{X}^{\top
}\Omega \mathbf{X}\left(  \mathbf{X}^{\top}\mathbf{X}\right)  ^{-1}$ to reflect the heteroscedasticity and dependence structure of the error vector. 
An important practical challenge in implementing such an approach is that the matrix $\Omega$ may be hard to estimate if the dependence structure of the error vector $\bm{\epsilon}$ is unknown. 
In this paper, we study the implications for the variance of OLS estimators of having a regressor of interest whose values are i.i.d.\ across units/time periods and are independent of values of other regressors.
The primary motivating applications for our analysis are randomized controlled trials in which units are independently assigned to some treatment level without a connection to observable characteristics. 
The main finding of the paper is that variance estimation in this case is often simplified, sometimes substantially.

Let $\mathbf{D}$ be the column of $\mathbf X$ corresponding to the regressor of interest and let $\mathbf{W}$ be the remaining columns of $\mathbf{X}$; i.e. columns corresponding to controls. Our first main result shows that when the vector $\mathbf D$ has i.i.d. components and is {\em strongly exogenous} in the sense of being independent not only of $\mathbf W$ but also of $\bm{\epsilon}$, the OLS estimator is asymptotically normally distributed and the formula $s^{2}\left(
\mathbf{X}^{\top}\mathbf{X}\right)  ^{-1}$ actually yields a valid variance estimator for the coefficient of $\mathbf D$ even if $\Omega$ is not proportional to the identity matrix. This result, which superficially contradicts the lessons of elementary linear regression analysis, is due to the randomness of
the $\mathbf{X}$ matrix in our analysis. While the textbook analysis assumes
away the randomness by conditioning on $\mathbf{X}$, we instead obtain our result by recognizing that the
randomness of the $\mathbf{X}$ matrix delivers a suitable martingale
structure.\footnote{The validity of the formula $s^{2}\left(  \mathbf{X}^{\top}%
\mathbf{X}\right)  ^{-1}$ does not mean that the variance estimators based on the formula $\left(  \mathbf{X}^{\top}\mathbf{X}\right)  ^{-1}\mathbf{X}^{\top
}\Omega \mathbf{X}\left(  \mathbf{X}^{\top}\mathbf{X}\right)  ^{-1}$
are invalid; see Lemma \ref{A_L2} in the Appendix for the asymptotic
equivalence of these estimators in our
setting.}\ We recognize that a version of this result in some simple contexts
is well understood in the profession in the sense that many can anticipate
such a result when the entire matrix of regressors is strongly exogenous; see references below. We
go one step further, however, and establish our result for the case when (i)
only \emph{one} regressor is strongly exogenous (e.g.,
treatment in a randomized controlled trial); and/or (ii) the error vector is
subject to some generalized dependence more complicated than what is commonly
understood to be the cluster structure, e.g. temporal/spatial autocorrelation
or a network structure. This result is important because it facilitates inference on the coefficient of the regressor of interest even if the researcher does not know the dependence structure of the error vector $\bm{\epsilon}$, which is useful from the pragmatic point of view. We emphasize, however, that while our conclusions hold
for the coefficient corresponding to a strongly exogenous regressor, they need
not hold for other coefficients in the regression.

Our second main result shows that when the vector $\mathbf D$ has i.i.d. components and is independent of $\mathbf W$ but $\bm{\epsilon}$ is {\em conditionally heteroscedastic} with respect to $\mathbf D$, the formula $s^2(\bf{X}^{\top}\bf{X})^{-1}$ is actually not valid and has to be adjusted not only for heteroscedasticity {\em but also}, rather surprisingly, for the dependence structure of the vector $\bm{\epsilon}$. Nevertheless, a simplified variance formula can often be used in this case as well. For example, conditional heteroscedasticity arises when the regression model $\mathbf{Y}=\mathbf{X}\theta+\bm{\epsilon}$ is taken from the potential outcomes framework with heterogeneous treatment effects.
If the researcher is concerned about clustering in this case, our results imply that it is sufficient to adjust the variance formula for clustering of the treatment effects only.\footnote{When the regression model $\mathbf{Y}=\mathbf{X}\theta+\bm{\epsilon}$ is taken from the potential outcomes framework with heterogeneous treatment effects, the case of strongly exogenous regressor corresponds to the assumption of constant treatment effects.} 
In contrast, for example, there is no need to adjust the variance formula for clustering of the potential outcomes in any given treatment arm. 
We also note that neither of our results restrict or exclude conditional heteroscedasticity of $\bm{\epsilon}$ with respect to $\mathbf W$.

In addition, we extend both results to the case when the values of the regressor of interest are independent only {\em across groups} of units/time periods, such as is the case in randomized controlled trials in which treatment assignment is
determined at a group (e.g., school/village) level. We show that in the strongly exogenous case, it suffices to take into account only the within-group correlation of the error vector $\bm{\epsilon}$. In other words, it suffices to use variance estimators that are clustered at the level at which treatment is assigned. In the conditional heteroscedasticity case, the variance formula still requires some adjustments for both heteroscedasticity and dependence but often allows for some simplifications relative to the standard textbook formula mentioned above.

Our first main result and its extension to the group-level assignment are related to but different from those in \cite{BDIK12}, who came to the same conclusions for regressions without controls and in which a fixed fraction of units/clusters is randomly assigned to be treated. To the best of our knowledge, however, there are no results in the literature related to our second main result.
Our analysis is also related to \cite{abadie2017should}, who presented a new clustering framework that emphasizes a finite population perspective as well as interactions between the sampling and assignment parts of the
data-generating process. They established in particular that there is no need to cluster when estimating the variance if the randomization is
conducted at the individual level and there is no heteregeneity in the treatment effects. Our first main result echoes and complements their findings in the following aspects. First, unlike them, we do not impose a particular
structure on the sampling process, which allows us to cover general forms of
time series or even network dependence in addition to the
cluster-type dependence. Second, our analysis goes beyond the binary treatment
framework and accommodates a general strongly exogenous regressor as well as
the inclusion of additional controls in the regression. In particular, we
allow for general dependence structures in the control variables, which makes
it ex-ante unclear at what level one should cluster. Third, we rely on a traditional asymptotic framework, which may
make our analysis more familiar to the reader. We recognize, however, that the
third aspect may be a weakness in the sense that our framework is unable to
address the finite population adjustment that plays an important role in
\cite{abadie2017should}.
Finally, we note \cite{bloom2005learning} who considered a random effects type cluster structure and produced a variance estimator for the simple difference of means estimator that is quoted in \cite{duflo2007using}. 
His equation (4.3), which is presented without proof, is in fact a special case of the $s^{2}\left(  \mathbf{X}^{\top}\mathbf{X}\right)  ^{-1}$ formula.
The cluster structure that he analyzed has a built-in correlation among observations, and as such, it would be tempting to think that variance estimation would require \cite{moulton1986random}'s clustering formula -- a conclusion that can be motivated if inference is to be conditioned on the $\mathbf{X}$ matrix. 
Hence, in our view, his equation\ (4.3) can only be motivated by explicitly recognizing the randomness of the $\mathbf{X}$ matrix.

Our results are not particularly difficult to derive. On the other hand, we are
unaware of any systematic discussion of results along this line in the
literature besides \cite{BDIK12} and \cite{abadie2017should}, especially in models where the
control variables and treatment variables are both present. Our results have
convenient pragmatic implications, which we hope are helpful to some empirical researchers.

\textit{Outline.} We present the basic intuition underlying our results in Section
\ref{sec:intuition}. The intuitive discussion there suggests that asymptotic
normality and the formula $s^{2}\left(  \mathbf{X}^{\top}\mathbf{X}\right)
^{-1}$ are valid under fairly general dependence structure in $\bm{\epsilon}$ provided that the randomness of $\mathbf{X}$ generates a suitable martingale structure in $\mathbf{X}^{\top}\bm{\epsilon}$. We also explain how conditional heteroscedasticity breaks down this martingale structure. We formalize our discussion in Section \ref{sec:main}, where our main
restriction on the dependence of $\bm{\epsilon}$ is that it be weak enough for its sample average to converge in probability
to zero -- a condition that further emphasizes that our analysis is driven by
the randomness in $\mathbf{X}$ and not $\bm{\epsilon}$. We provide an extension to the case of group-level random assignment in Section \ref{sec:group}.

%We also emphasize that our result in Section \ref{sec:main} is applicable
%even when a researcher does not know the precise correlation/cluster
%structure, which is arguably important from a pragmatic point of view.\ On the
%other hand, our conclusions can fail to hold if the martingale structure in
%$\mathbf{X}^{\top}\bm{\epsilon}$ breaks down. Two empirically relevant cases when this happens are discussed
%in details: (i) The error vector $\bm{\epsilon}$ is conditional heteroscedastic with respect to $\mathbf{X}$, such as is the
%case in randomized controlled trials with heterogeneous treatment effects; and
%(ii) The coordinates of $\mathbf{X}$ exhibit a group structure, such as is the
%case in randomized controlled trials in which treatment assignment is
%determined at a  group (e.g., school/village) level. We examine these
%challenges in Sections \ref{sec:main} and \ref{sec:group}.

\textit{Notation.} We use $K$ to denote a generic strictly positive constant
that may change from place to place but is independent of the sample size $n$. For any positive integer $k$, let
$\mathbf{I}_{k}$, $\mathbf{1}_{k\times1}$, and $\mathbf{0}_{k\times 1}$ denote the $k\times k$ identity
matrix, $k\times1$ vector of ones, and $k\times1$ vector of zeros. For any real square matrix $A$,
$\lambda_{\min}(A)$ and $\lambda_{\max}(A)$\ denote its smallest and largest
eigenvalues. We use $A\equiv B$ to denote that $A$ is defined as $B$, and
$(A_{j})_{j\leq J}$ to denote the matrix composed by sequentially stacking
matrices $A_{1},\ldots,A_{J}$ with equal number of columns.\
%Throughout this paper, we write $\sum_{i}$ for $\sum_{t\leq n}$ and $\sum_{j,i}$ for the double summation $\sum_{j\leq n_{g}}\sum_{i\leq n_{j}}$, where\ $n_{j}$ and $n_{g}$ denote the size of group $j$ and the number of groups respectively, and $n$ denotes the sample size such that $n=\sum_{j\leq n_{g}}n_{j}$.

\section{Intuition\label{sec:intuition}}

In this section, we provide intuition for the validity of the formula $s^{2}\left(
\mathbf{X}^{\top}\mathbf{X}\right)  ^{-1}$ in the case of strongly
exogenous regressors. We first consider the case of a simple univariate
regression model and then extend the result to the case of a multivariate regression model. At the end of this section, we explain complications arising conditional heteroscedasticity and how they break the validity of the formula $s^{2}\left(
\mathbf{X}^{\top}\mathbf{X}\right)  ^{-1}$.

\subsection{Case of Strong Exogeneity}

%\subsection{Univariate Regression Model}

We start by considering a simple univariate linear regression time series
model in which we have
\begin{equation}
\label{simple-regression}y_{i}=d_{i}\beta+\varepsilon_{i},\quad i=1,\ldots,n,
\end{equation}
where $(\varepsilon_{i})_{i\leq n}$ is a second-order, possibly
autocorrelated, stationary time series -- we employ the index $i$, rather than
$t$, to emphasize our analysis is not confined to the time series context. We
depart from the textbook time series model by assuming that the regressors
$(d_{i})_{i\leq n}$ are: (i) Independent and identically distributed (i.i.d.)
with mean zero, and (ii) Strongly exogenous in the sense that $(d_{i})_{i\leq
n}$ is independent of the time series process $(\varepsilon_{i})_{i\leq n}$.

As is well-known, the least squares estimator $\hat{\beta}$ of $\beta$ in
model \eqref{simple-regression} satisfies the equality
\begin{equation}
\label{simple-expansion}\sqrt{n}(\hat{\beta}-\beta)=\frac{n^{-1/2}\sum
_{i=1}^{n}d_{i}\varepsilon_{i}}{n^{-1}\sum_{i=1}^{n}d_{i}^{2}}.
\end{equation}
In many standard time series textbooks, the asymptotic distribution of
$\hat \beta$ is thus derived by imposing sufficiently strong conditions to
ensure that the score $n^{-1/2}\sum_{i=1}^{n}d_{i}\varepsilon_{i}$ is
asymptotically normal and the Hessian $n^{-1}\sum_{i=1}^{n}d_{i}^{2}$
converges in probability to a non-stochastic matrix. In order to derive the
standard error for $\hat{\beta}$, we therefore only need a consistent
estimator of the long-run variance of the score; i.e., a heteroscedasticity
and autocorrelation consistent (HAC) variance estimator, such as those
proposed by \cite{neweywest1987} and \cite{andrews1991hac}.

On the other hand, if the regressors $(d_{i})_{i\leq n}$ are i.i.d.\ and
strongly exogenous with mean zero, the independence of $(d_{i})_{i\leq
n}$ and $(\varepsilon_{i})_{i\leq n}$ implies that for any $1\leq i_1,i_2\leq n$ with $i_{1}\neq i_{2}$,
we must have
\[
\mathbb{E}\left[  d_{i_{1}}\varepsilon_{i_{1}}d_{i_{2}}\varepsilon_{i_{2}%
}\right]  =\mathbb{E}\left[  d_{i_{1}}\right]  \mathbb{E}\left[  d_{i_{2}%
}\right]  \mathbb{E}\left[  \varepsilon_{i_{1}}\varepsilon_{i_{2}}\right]
=0,
\]
and also $\mathbb{E}[(d_{i}\varepsilon_{i})^{2}]=\mathbb{E}[d_{i}%
^{2}]\mathbb{E}[\varepsilon_{i}^{2}]$ for all $1\leq i\leq n$. Hence, as long
as some version of the central limit theorem is applicable to the score
$n^{-1/2}\sum_{i=1}^{n}d_{i}\varepsilon_{i}$ and a law of large numbers is
applicable to the Hessian $n^{-1}\sum_{i=1}^{n}d_{i}^{2}$, we can conclude
that the asymptotic distribution of $\hat{\beta}$ is given by
\[
\sqrt{n}(\hat{\beta}-\beta)\rightarrow_{d}N\left(  0,\text{ \ }\frac
{\mathbb{E}\left[  \varepsilon_{i}^{2}\right]  }{\mathbb{E}\left[  d_{i}%
^{2}\right]  }\right)  .
\]
In particular, statistical inference on $\beta$ can be conducted as if it were
not a time series model,
i.e.\ using the formula $s^{2}\left(  \mathbf{X}^{\top}\mathbf{X}\right)
^{-1}$, where $\mathbf X\equiv (d_i)_{i\leq n}$ in this case.

The main takeaway of the preceding example is that the strong exogeneity and
i.i.d.\ nature of the regressors $(d_{i})_{i\leq n}$ imply that the sequence
$(d_{i}\varepsilon_{i})_{i\leq n}$ is homoscedastic \emph{even if} the errors
$(\varepsilon_{i})_{i\leq n}$ are arbitrarily autocorrelated. Is this
simplification confined to the time series model? Our preceding discussion
suggests that this is not the case. Indeed, provided the regressors
$(d_{i})_{i\leq n}$ are i.i.d., mean zero, and strongly exogenous, the score
$n^{-1/2}\sum_{i=1}^{n}d_{i}\varepsilon_{i}$ has a built-in martingale
structure vis-\`{a}-vis the filtration $\mathcal{F}_{i}$ generated by $(d_{j}
,\varepsilon_{j})_{j\leq i}^{\top}$ because:
\begin{equation}
\mathbb{E}\left[  \left.  d_{i}\varepsilon_{i}\right \vert \mathcal{F}%
_{i-1}\right]  =\mathbb{E}\left[  \left.  \mathbb{E}\left[  \left.
d_{i}\right \vert \mathcal{F}_{i-1},\varepsilon_{i}\right]  \varepsilon
_{i}\right \vert \mathcal{F}_{i-1}\right]  = \mathbb{E}[d_{i}]\mathbb{E}%
[\varepsilon_{i}\vert \mathcal{F}_{i-1}] =0 . \label{mds_1}%
\end{equation}
Therefore, assuming that the random pairs $(d_{i},\varepsilon_{i})$
satisfy certain moment conditions, the martingale central limit theorem will
be applicable regardless of the dependence structure of $(\varepsilon
_{i})_{i\leq n}$ and the long run variance of the score will reduce to
$\mathbb{E}[d_{i}^{2}]\mathbb{E}[\varepsilon_{i}^{2}]$. In particular, the
variance formula $s^{2}\left(  \mathbf{X}^{\top}\mathbf{X}%
\right)  ^{-1}$ will remain valid despite the dependence present in the
variables $(\varepsilon_{i})_{i\leq n}$. Thus, spatial correlation, network
dependence, and/or a cluster structure in the variables $(\varepsilon
_{i})_{i\leq n}$ are all accommodated by the standard homoscedastic standard
errors. Moreover, we note that a quick inspection at the preceding argument
reveals that the assumptions we have imposed so far are stronger than
necessary for the desired conclusion to hold.
%\footnote{Observe also that the assumption of stationarity that we imposed on $(\varepsilon_t)_{t\leq n}$ was used to simplify exposition but played no role in our derivation: without it, we would simply have to replace $\mathbb[\varepsilon_t^2]$ in the asymptotic variance formula by more cumbersome $n^{-1}\sum_t \mathbb E[\varepsilon_t^2]$, without changing the conclusion of the applicability of the $s^{2}\left(\mathbf{X}^{\top}\mathbf{X}\right)  ^{-1}$ formula.}

%\subsection{Multivariate Regression Model}

We next build on our preceding discussion by considering the multivariate
linear regression model
\begin{equation}
y_{i}=d_{i}\beta+w_{i}^{\top}\gamma+\varepsilon_{i}, \quad i=1,\ldots,n,
\label{multiple-regression}%
\end{equation}
where $d_i$ is a scalar regressor of interest, $w_{i}$ is a $d_w$-vector of controls, and $(w_{i},\varepsilon_{i})_{i\leq n}$ is a
second-order, possibly autocorrelated, stationary time series satisfying
$\mathbb{E}[w_{i}\varepsilon_{i}] = \mathbf{0}_{d_w\times 1}$ for all $1\leq i\leq n$. We continue to assume that the regressors $(d_{i})_{i\leq n}$ are
i.i.d.\ with mean zero, and strongly exogenous in the sense that
$(d_{i})_{i\leq n}$ is independent of the time series process $(\varepsilon_{i},w_{i}^{\top})_{i\leq n}$. The parameter of interest continues to be
$\beta$.

For this model, the Frisch-Waugh-Lovell theorem implies the least squares
estimator $\hat \beta$ satisfies
\begin{equation}
\label{OLS-M}\sqrt{n}\left(  \hat \beta- \beta \right)  = \frac{n^{-1/2}
\sum_{i=1}^{n}\left(  d_{i} - w_i^{\top}\hat \alpha \right)  \varepsilon_{i}%
}{n^{-1}\sum_{i=1}^{n} \left(  d_{i} - w_i^{\top}\hat \alpha \right)  ^{2}}
\hspace{0.5 in} \hat \alpha \equiv \left(\sum_{i=1}^n w_iw_i^{\top}\right)^{-1}\sum_{i=1}^n w_id_i.
\end{equation}
Hence, under appropriate regularity conditions the estimator $\hat \beta$
admits the asymptotic expansion
\begin{equation}
\label{OLS-FWL}\sqrt{n}\left(  \hat \beta- \beta \right)  = \frac{n^{-1/2}
\sum_{i=1}^{n}\left(  d_{i} - w_i^{\top}\alpha  \right)  \varepsilon_{i}}%
{n^{-1}\sum_{i=1}^{n} \left(  d_{i} - w_i^{\top}\alpha\right)  ^{2}} + o_{p}(1)
\hspace{0.5 in} \alpha \equiv (\mathbb E[w_iw_i^{\top}])^{-1}\mathbb E[w_id_i].
\end{equation}
In particular, if $(d_{i})_{i\leq n}$ is mean zero and independent of
$(w_{i}^\top)_{i\leq n}$, then $\alpha= 0$ and the asymptotic expansion reduces to the univariate setting -- i.e.\ the right-hand side of  \eqref{OLS-FWL} is
(asymptotically) equivalent to the right-hand side of \eqref{simple-expansion}.
It therefore follows that the end result is the same as in the univariate case: The variance
formula $s^{2}\left(  \mathbf{X}^{\top}\mathbf{X}\right)  ^{-1}$, where $\mathbf{X}\equiv (d_i,w_i^\top)_{i\leq n}$ in this case, remains valid
for $\hat \beta$. We emphasize, however, that this formula is
not necessarily justified for conducting inference on the coefficient $\gamma$.

As a preview of results in the next section, we again note that the conditions
we have imposed so far are stronger than required. For instance, suppose that
instead of demanding that the regressors $(d_{i})_{i\leq n}$ and
$(w_{i}^{\top})_{i\leq n}$ be fully independent, we assume that they are related
according to the model
\[
d_{i} = \alpha w_{i} + \eta_{i},%
\]
for $(\eta_{i})_{i\leq n}$ i.i.d.\ and independent of $(\varepsilon
_{i},w_i^{\top})_{i\leq n}$. The asymptotic expansion in \eqref{OLS-FWL}, combined with
the same arguments employed in the univariate case, then continue to imply
that
\[
\sqrt{n}(\hat{\beta}-\beta)\rightarrow_{d}N\left(  0,\text{ \ }\frac
{\mathbb{E}\left[  \varepsilon_{i}^{2}\right]  }{\mathbb{E}\left[
\eta_i^{2}\right]  }\right)
\]
under mild moment conditions -- i.e.\ when computing standard errors for
$\hat \beta$ we can continue to pretend that the time series process fits the
textbook homoscedastic model. Setting $w_{i}$ to be a constant, for instance,
reveals that the mean zero assumption on $(d_{i})_{i\leq n}$ is superfluous.

\subsection{Case of Conditional Heteroscedasticity}
%\subsection{Complications and Limitations}

The preceding martingale argument relies crucially on two key assumptions: (i)
The regressors $(d_{i})_{i\leq n}$ are independent of each other, and (ii) The
regressors $(d_{i})_{i\leq n}$ are independent of the errors $(\varepsilon
_{i})_{i\leq n}$. A challenge to our martingale argument arises when $(d_{i})_{i\leq
n}$ and $(\varepsilon_{i})_{i\leq n}$ are not independent. Within the potential outcome framework, for instance, this full independence requirement is violated in the presence of heterogenous treatment effects. More precisely,
heterogenous treatment effects render $\varepsilon_{i}$ conditionally
heteroscedastic with respect to the treatment status $d_{i}$. Motivated by
this observation, we also study a model in which $\varepsilon_{i} =\sum_{l=1}^L \sigma_{l}(d_{i})\varepsilon_{l,i}^{*}$ with $e_{i} \equiv(\varepsilon
_{l,i}^{*})_{l\leq L}$ possibly correlated across $i$ and $l$, but
$(e_{i})_{i\leq n}$ fully independent of $(d_{i})_{i \leq n}$. In the potential outcome framework, with $d_{i}\in\{0,1\}$ indicating treatment status, we would have
\begin{equation}
\label{eq:heteroex}\varepsilon_{i} = (y_{i}(0) -\mathbb{E}[y_{i}(0)]) +
d_{i}((y_{i}(1)-y_{i}(0))-\mathbb{E}[y_{i}(1)-y_{i}(0)]),
\end{equation}
where $y_{i}(d)$ denotes the potential outcome for unit $i$ under treatment
status $d\in\{0,1\}$.

To see the problem for the martingale structure in this model, observe that for any $1\leq i_1,i_2 \leq n$, we now have
\begin{align*}
\mathbb E[d_{i_1}\varepsilon_{i_1}d_{i_2}\varepsilon_{i_2}]
& = \mathbb E\left[d_{i_1}\sum_{l_1=1}^{L}\sigma_{l_1}(d_{i_1})\varepsilon_{l_1,i_1}^* d_{i_2}\sum_{l_2=1}^{L}\sigma_{l_2}(d_{i_2})\varepsilon_{l_2,i_2}^* \right] \\
& = \sum_{l_1=1}^{L}\sum_{l_2=1}^{L}\mathbb E\left[ d_{i_1}\sigma_{l_1}(d_{i_1}) \right]\mathbb E\left[ d_{i_2}\sigma_{l_2}(d_{i_2}) \right]\mathbb E[\varepsilon_{l_1,i_1}^*\varepsilon_{l_2,i_2}^*],
\end{align*}
which is not necessarily zero even if $d_i$'s are mean zero. The variance formula  for the sum $\sum_{i\leq n}d_i\varepsilon_i$ therefore must include interactions terms as long as the random vectors $(\varepsilon_{l,i_1}^*)_{l\leq L}$ and $(\varepsilon_{l,i_2}^*)_{l\leq L}$ are correlated.

We will present a detailed analysis of the conditional heteroscedasticity case in Section
\ref{sec:main} but the main takeaways from our results are: (i) The independence
of $(d_{i})_{i\leq n}$ from controls $(w_{i}^{\top})_{i\leq n}$ still
simplifies the asymptotic variance for $\hat \beta$; and (ii)
Conditional heteroscedasticity yields a break in the martingale structure
that requires us to adjust standard errors \emph{not only} for
heteroscedasticity, but \emph{also} for correlation of the errors across units.
For example, in the context of \eqref{eq:heteroex}, our analysis implies the
asymptotic variance of $\sqrt n(\hat \beta-\beta)$ equals (the probability limit of)
\begin{equation}\label{eq: variance treatment effect formula}
\frac{1}{n\sigma_{d}^{4}}\sum_{i=1}^{n} \mathbb{E}[(d_{i} - \mathbb{E}%
[d_{i}])^{2} \varepsilon_{i}^{2}] +
\text{Var}\left( n^{-1/2} \sum_{i=1}^{n} (y_{i}(1)-y_{i}(0))\right)  - n^{-1}\sum_{i=1}^{n}
\text{Var}\left(  y_{i}(1)-y_{i}(0)\right)   .
\end{equation}
In particular, we note that standard errors may need to be adjusted for correlation if we are concerned the treatment effects are correlated. On the other hand, the correlation between components of the vector $(y_i(0))_{i\leq n}$ plays no role for the standard errors, and neither does correlation between components of the vector $(y_i(1))_{i\leq n}$. %Combining with our preceding discussion on group level randomization, for instance, it follows that in order for clustering at the level at which treatment is assigned to be justified we must have that treatment effects are uncorrelated across treatment groups.

\section{Main Theory\label{sec:main}}

We now present a rigorous theory showing that the variance formula $s^{2}\left(\mathbf{X}^{\top}\mathbf{X}\right)^{-1}$ for the OLS estimator is valid in the strongly exogenous case even if the errors in the regression model are correlated. We also derive the variance formula in the case of conditional heteroscedasticity. In what follows, we suppose an outcome $y_{i}$ satisfies
\begin{equation}\label{eq: regression model null}
y_i = x_i^{\top}\theta + \varepsilon_i,\quad i=1,\ldots,n,
\end{equation}
where $x_i = (d_i,w_i^\top)^\top$ is a vector of regressors, with $d_i$ being a key regressor and $w_i$ being a vector of controls, $\theta = (\beta,\gamma^\top)^\top$ is a vector of parameters, with $\beta$ being a parameter of interest and $\gamma$ being a vector of nuisance parameters, and $\varepsilon_i$ is an error term with mean zero. More explicitly, the regression model \eqref{eq: regression model null} can be rewritten as
\begin{equation}
y_{i}=d_{i}\beta+w_{i}^{\top}\gamma+\varepsilon_{i},\quad i=1,\ldots,n.
\label{M1}%
\end{equation}
We assume that the first component of the vector $w_i$ is a (non-zero) constant, meaning that the regression model \eqref{M1} contains the intercept term.  For notational simplicity, we set $\mathbf{Y}\equiv(y_{i})_{i\leq n}$, $\mathbf{D}\equiv(d_{i})_{i\leq n}$,
$\mathbf{W}\equiv(w_{i}^{\top})_{i\leq n}$, $\mathbf{X} \equiv (x_i^{\top})_{i\leq n}$, and $\bm{\epsilon}\equiv(\varepsilon_{i})_{i\leq n}$. %We also set $u_{i}\equiv(d_{i},w_{i}^{\top})^{\top}\varepsilon_{i}$.

Under a suitable
exogeneity assumption on $(d_{i},w_{i}^{\top})^{\top}$, the unknown parameter
$\theta \equiv(\beta,\gamma^{\top})^{\top}$ can be estimated by OLS:
$$
\hat{\theta}\equiv(\hat{\beta},\hat{\gamma}^{\top})^{\top}\equiv (\mathbf X^{\top}\mathbf X)^{-1}(\mathbf X^{\top}\mathbf Y).
$$
Standard estimators of the asymptotic variance of $\hat{\theta}$ rely on the
asymptotic variance of the ``score" $n^{-1/2}\sum_{i=1}^{n} (d_{i},w_{i}^{\top})^{\top}\varepsilon_{i}$, which may
take a complicated form due to possible dependence between observations. This relationship makes standard error estimation and
statistical inference challenging in practice. For instance,  cluster-robust standard errors, such as those proposed by \cite{moulton1986random}, \cite{LiangZeger86},
and \cite{arellano1987computing}, are predicated on knowledge of the relevant
group structure at which to cluster; see \cite{hansen2007asymptotic},
\cite{ibragimov2010t} and \cite{ibragimov2016inference} for related
discussion. Similarly, spatial standard errors, such as those proposed by
\cite{conley1999gmm}, often require knowledge of a measure of ``economic
distance" that is relates to the degree of dependence across observations.
However, we next show that when $\mathbf{D}$ is independent of $\mathbf{W}$,
such as in randomized control trials, the asymptotic variance for $\hat \beta$
simplifies significantly. As a result, estimation of standard errors and
asymptotically valid inference simplify as well.

\subsection{Case of Strong Exogeneity}

We first study the case that $\mathbf{D}$ is independent of $%
%TCIMACRO{\TeXButton{epsilon}{\bm{\epsilon}}}%
%BeginExpansion
\bm{\epsilon}%
%EndExpansion
$, and hence $\bm{\epsilon}$ is conditionally homoscedastic with
respect to $\mathbf{D}$. Together with independence of $\mathbf D$ from $\mathbf W$, this means that $\mathbf D$ is strongly exogenous. The conditional heteroscedasticity case is discussed
later in this section. In the assumptions that follow, recall that $K$
should be interpreted to denote a sufficiently large constant that can change from place to place but is independent of the sample size $n$.

\begin{assumption}
\label{A1}(i)\ The random variables $d_{i}$, $1\leq i\leq n$, are i.i.d. with mean $\mu_d$ and variance $\sigma_d^2$; (ii) $(d_{i})_{i\leq
n}$ is independent of $(w_i^{\top})_{i\leq n}$;  (iii)\ there exists a constant $\delta_1>0$ such that
$\max_{i\leq n}\mathbb{E}[|d_{i}|^{2+\delta_1}]\leq K$; (iv) $\sigma_{d}^{2}\geq K^{-1}$; (v) $(d_{i})_{i\leq
n}$ is independent of $(\varepsilon_i)_{i\leq n}$.
\end{assumption}

\begin{assumption}
\label{A2}(i) $\lambda_{\min}(n^{-1}
\sum_{i=1}^{n}w_i w_i^{\top})\geq K^{-1} + o_p(1)$; (ii)
$\max_{i\leq n} \mathbb{E}[||w_{i}||^{2}]\leq K$; (iii) the first component of the vectors $w_i$, $1\leq i\leq n$, is a non-zero constant.
\end{assumption}

\begin{assumption}
\label{A3}(i)\ $n^{-1}\sum_{i=1}^{n}w_{i}\varepsilon_{i}=o_{p}(1)$; (ii)
$n^{-1}\sum_{i=1}^{n}\varepsilon_{i}^{2}=n^{-1}\sum_{i=1}^{n}\mathbb{E}%
[\varepsilon_{i}^{2}]+o_{p}(1)$; (iii) there exists a constant $\delta_2>0$ such that $\max_{i\leq n}\mathbb{E}%
[|\varepsilon_{i}|^{2+\delta_2}]\leq K$; (iv) $n^{-1}\sum_{i=1}^{n}\mathbb{E}%
[\varepsilon_{i}^{2}]\geq K^{-1}$.
\end{assumption}

Assumption \ref{A1} contains our main requirements for the regressor of interest. In particular, Assumptions \ref{A1}(i) and \ref{A1}(ii) are our key conditions that are satisfied in many randomized control trials. Assumptions \ref{A1}(iii) and \ref{A1}(iv) are mild moment conditions. Assumption \ref{A1}(v) means that $(d_{i})_{i\leq n}$ is strongly exogenous. Assumption \ref{A2} contains our main requirements for the controls. In particular, Assumption \ref{A2}(i) means that there is no multicollinearity among controls. Assumption \ref{A2}(ii) is a mild moment condition. Assumption  \ref{A2}(iii) means that we study regressions with an intercept. Assumption \ref{A3} contains our main requirements for the regression error. Assumption \ref{A3}(i) holds if $\varepsilon_i$'s are uncorrelated with $w_i$'s and a law of large numbers applies to the product $w_i\varepsilon_i$. Assumption \ref{A3}(ii) is essentially a law of large numbers for $\varepsilon_i^2$'s. Assumptions \ref{A3}(iii) and \ref{A3}(iv) are mild moment conditions. We highlight that our assumptions allow for a wide array of dependence
structures in the matrix $(\varepsilon_{i},w_{i}^{\top})_{i\leq n}$, with the main
condition in this regard intuitively being that dependence be
\textquotedblleft weak" enough for the law of large numbers imposed in
Assumptions \ref{A3}(i) and \ref{A3}(ii) to apply.

For all $1\leq i\leq n$, denote $d_i^*\equiv d_i - \mu_d$. The following theorem derives the asymptotic distribution of the OLS estimator $\hat{\beta}$ in the strongly exogenous case.

\begin{theorem}
\label{L1}Let Assumptions \ref{A1}, \ref{A2} and \ref{A3} hold. Then
\[
\frac{\sqrt{n}(\hat{\beta}-\beta)}{\sigma_{\varepsilon}/\sigma_{d}} = \frac{n^{-1/2}\sum_{i=1}^n d_i^*\varepsilon_i}{\sigma_d\sigma_{\varepsilon}} + o_P(1) \rightarrow_{d}N(0,1),
\]
where $\sigma_{\varepsilon}^{2}\equiv n^{-1}\sum_{i=1}^{n}\operatorname{Var}%
(\varepsilon_{i})$.
\end{theorem}

This theorem establishes two key facts. First, it shows that, given the
strong exogeneity of $(d_{i})_{i\leq n}$, our mild requirements on the
dependence structure of $(\varepsilon_{i},w_{i}^{\top})_{i\leq n}$ suffice for
establishing asymptotic normality of $\hat{\beta}$. To establish such a
conclusion, we rely on a martingale construction that generalizes our
discussion in Section \ref{sec:intuition}. Second, Theorem \ref{L1} establishes
that the asymptotic variance of $\hat{\beta}$ is not affected by the possible correlation across vectors $w_i\varepsilon_{i}$ since it
only depends on the variance of $d_{i}$ and the averaged variance of the error
terms $(\varepsilon_{i})_{i\leq n}$. We emphasize that neither of these
conclusions need hold for the estimator $\hat{\gamma}$ of the coefficient
$\gamma$ corresponding to the vectors of controls $w_i$.%

In addition, Theorem \ref{L1} suggests that we can estimate the variance of $\hat{\beta}$ by
$s^{2}(\mathbf{\breve{D}^{\top}\breve{D}})^{-1}$, where
\begin{equation}
s^{2}\equiv n^{-1}(\mathbf{Y}-\mathbf{D}\hat{\beta
}-\mathbf{W}\hat{\gamma})^{\top}(\mathbf{Y}-\mathbf{D}\hat{\beta}%
-\mathbf{W}\hat{\gamma}), \label{var_est}%
\end{equation}
$\mathbf{\breve{D}}\equiv \mathbf{M}_{W}\mathbf{D}$ and $\mathbf{M}_{W}\equiv \mathbf{I}%
_{n}-\mathbf{W}(\mathbf{W}^{\top}\mathbf{W})^{-1}\mathbf{W}^{\top}$. The following
corollary confirms this conjecture.

\begin{corollary}
\label{T1}Let Assumptions \ref{A1}, \ref{A2} and \ref{A3} hold. Then
\begin{equation}
\label{eq: asy variance estimator consistency}
s^{2}(n^{-1}\mathbf{\breve{D}^{\top}\breve{D}%
})^{-1}=\sigma_{\varepsilon} ^{2}/\sigma_{d}^{2}+o_{p}(1)
\text{ \ \ \ and  \ \ \ }
\frac
{\hat{\beta}-\beta}{\sqrt{s^{2}(\mathbf{\breve{D}^{\top}\breve{D}})^{-1} }%
}\rightarrow_{d}N(0,1).
\end{equation}
\end{corollary}
Together with Theorem \ref{L1}, this corollary is our first main result. Indeed, it is well-known that the top left element of the matrix $(\mathbf{X}^{\top
}\mathbf{X})^{-1}$ coincides with
$(\mathbf{\breve D}^{\top}\mathbf{\breve D})^{-1}$.
Therefore, this corollary justifies using the
classic standard error formula $s^{2}\left(
\mathbf{X}^{\top}\mathbf{X}\right)  ^{-1}$ for inference on $\beta$, even though we allow for general
dependence structures in the errors $(\varepsilon_{i})_{i\leq n}$ and
controls $(w_{i}^\top)_{i\leq n}$.

\bigskip

\noindent \textsc{Remark}. Theorem \ref{L1} and Corollary \ref{T1} can be extended to allow for
dependence between $d_{i}$ and $w_{i}$. Indeed, suppose that $d_{i}$ depends linearly on $w_{i}$:
\[
d_{i}=w_{i}^{\top}\alpha+\eta_{i},
\]
where $\eta_{i}$'s satisfy the conditions of Assumption \ref{A1} imposed on
$d_{i}$. By arguments that are similar to those in the proof of Theorem \ref{L1} and Corollary \ref{L1}, it
is then straightforward to show that
\[
\frac{\sqrt{n}(\hat{\beta}-\beta)}{\sigma_{\varepsilon}/\sigma_{\eta}}\rightarrow_{d}N(0,1),
\]
where $\sigma_{\eta}^{2}$ denotes the variance of $\eta_i$'s, and that the convergence results \eqref{eq: asy variance estimator consistency} still hold. %Moreover,
%$\sigma_{\varepsilon}^{2}$ can be estimated by the same estimator $s^{2}$ in
%(\ref{var_est}), and $\sigma_{\eta}^{2}$ can be estimated by $n^{-1}\sum
%_{i=1}^{n}\hat{\eta}_{i}^{2}$ for $\hat{\eta}_{i}$ residual from regressing
%$(x_{i})_{i\leq n}$ on $(w_{i}^{\top})_{i\leq n}$.
\qed

\bigskip

\noindent
\textsc{Remark}. Theorem \ref{L1} and Corollary \ref{T1} can also be extended to instrumental variable (IV) estimators. Indeed, suppose that
\begin{equation}\label{eq: first stage}
d_{i}=v_{i}\rho+w_{i}^{\top}\alpha+\eta_{i},
\end{equation}
where $v_i$ is an instrumental variable satisfying the conditions of Assumption \ref{A1} imposed on $d_i$ and $\eta_i$ is a (first-stage) estimation error satisfying the conditions of Assumption \ref{A3} imposed on $\varepsilon_i$. 
In addition denote $\mathbf{V}\equiv(v_{i})_{i\leq n}$ and define $\mathbf{M}_{V,W}$
and $\mathbf{M}_{V}$ the same way as $\mathbf{M}_{W}$ with $\mathbf{W}$
replaced by $\left(  \mathbf{V},\mathbf{W}\right)  $ and $\mathbf{V}$,
respectively. The two-stage least squared (2SLS) estimator of $\beta$ then
satisfies%
\[
\hat{\beta}_{2sls}=\frac{\mathbf{D}^{\top}(\mathbf{M}_{W}-\mathbf{M}%
_{V,W})\mathbf{Y}}{\mathbf{D}^{\top}(\mathbf{M}_{W}-\mathbf{M}_{V,W}%
)\mathbf{D}}.
\]
Using Assumptions \ref{A2} and \ref{A3}, it is then possible to show that
\[
\sqrt{n}(\hat{\beta}_{2sls}-\beta)=\frac{n^{-1/2}\sum_{i=1}^{n}(v_{i}-\mu_v)\varepsilon_{i}}{\rho\sigma_{v}^{2}}+o_{p}(1),
\]
where $\mu_v$ is the mean of $v_i$'s and $\sigma_v^2$ is the variance of $v_i$'s. Therefore,
\[
\frac{\sqrt{n}(\hat{\beta}_{2sls}-\beta)}{\sigma_{\varepsilon}/(\rho \sigma
_{v})}\rightarrow_{d}N(0,1).
\]
It is clear that $\sigma_{\varepsilon}^{2}$ can be estimated by the same
$s^{2}$ as that in (\ref{var_est}) with $\hat{\beta}$ replaced by
$\hat{\beta}_{2sls}$, $\rho$ can be estimated by OLS on the (first-stage) regression \eqref{eq: first stage}, and $\sigma_v^2$ can be estimated by $n^{-1}\mathbf{\tilde{V}^{\top}\tilde{V}}$, where $\mathbf{\tilde V} = \mathbf V - \mathbf{1}_{n,1}(n^{-1}\mathbf V^\top\mathbf 1_{n\times1})$. \qed

\subsection{Case of Conditional Heteroscedasticity}

Next, we derive the asymptotic variance for $\hat
{\beta}$ in the case of conditional heteroscedasticity, i.e. when $(\varepsilon_{i})_{i\leq n}$ is conditionally heteroscedastic
with respect to $(d_{i})_{i\leq n}$. Following the notation introduced in
Section \ref{sec:intuition}, we focus on the case in which $\varepsilon
_{i}\equiv \sum_{l\leq L}\sigma_{l}(d_{i})\varepsilon_{l,i}^{\ast}$, where
$e_{i}\equiv(\varepsilon_{l,i}^{\ast})_{l\leq L}$ is a vector with mean zero, for all $1\leq i \leq n$.

Let $A_{i}\equiv (d_{i} - \mu_d)(\sigma_{l}(d_{i}))_{l\leq L}$ for all $1\leq i\leq n$ and observe that under Assumption \ref{A1}(i), the random vectors $A_i$ are i.i.d. Denote their common mean vector by $\mu_A$.  In addition, denote $\sigma_{e,1}^2\equiv n^{-1}\sum_{i=1}^n \mathbb E[((A_i - \mu_A)^{\top}e_i)^2]$ and $\sigma_{e,2}^2\equiv \mathbb E[(n^{-1/2}\sum_{i=1}^n\mu_A^{\top}e_i)^2]$. Within this context, we impose the following assumptions.

\begin{assumption}
\label{A4}(i) $(d_{i})_{i\leq n}$ is independent of $(e_{i})_{i\leq n}$; (ii) the functions $\sigma_l$, $1\leq l\leq L$, are bounded; (iii) $\sigma_{e,1}^2 \geq K^{-1}$.
\end{assumption}
\begin{assumption}
\label{A5new} (i) $\|n^{-1}\sum_{i=1}^n e_ie_i^{\top} - n^{-1}\sum_{i=1}^n\mathbb E[e_ie_i^{\top}]\| = o_p(1)$; (ii) there exists a constant $\delta_3 >0$ such that $n^{-1}\sum_{i=1}^n\|e_i\|^{2+\delta_3} = O_p(1)$; (iii) $\sigma_{e,2}^{-1}n^{-1/2}\sum_{i=1}^n\mu_A^{\top}e_i \to_d N(0,1)$.
\end{assumption}

Assumption \ref{A4} is mainly used to replace Assumption \ref{A1}(v) and
accounts for the conditional heteroscedasticity of $(\varepsilon_{i})_{i\leq
n}$ with respect to $(d_i)_{i\leq n}$. Assumption \ref{A4}(i) requires that $(d_{i})_{i\leq n}$ are strongly
exogenous with respect to the``scaled" error vector $(e_{i})_{i\leq n}$ -- a
requirement that, as discussed in Section \ref{sec:intuition} maps well into a potential outcome framework with heterogeneous treatment effects. Assumption
\ref{A4}(ii) imposes upper bounds on the functions $\sigma_{l}^{2}(\cdot)$, which we view as
a mild regularity condition. Assumption \ref{A4}(iii) is a mild moment condition. Assumption \ref{A5new} contains further restrictions on the vectors $e_i$. Assumption \ref{A5new}(i) is essentially a law of large numbers for $e_ie_i^\top$'s. Assumption \ref{A5new}(ii) is essentially a moment condition for the random variables $\|e_i\|$. Assumption \ref{A5new}(iii) limits the amount of dependence among the vectors $e_i$ to ensure convergence in distribution.

The following theorem, which is our second main result, derives the asymptotic distribution of the OLS estimator $\hat\beta$ in the case of conditional heteroscedasticity.

\begin{theorem}
\label{L3}Let Assumptions \ref{A1}(i)-(iv), \ref{A2}, \ref{A3},
\ref{A4}, and \ref{A5new} hold. Then
\begin{equation}
\frac{\sqrt{n}(\hat{\beta}-\beta)}{\sigma_{d\varepsilon}/\sigma_d^2}=\frac{n^{-1/2}\sum_{i=1}^{n}d_{i}^*\varepsilon_{i}}{\sigma_{d\varepsilon}}+o_{p}(1) \to_d N(0,1), \label{linear-exp}%
\end{equation}
where $\sigma_{d\varepsilon}^2 \equiv \sigma_{e,1}^2 + \sigma_{e,2}^2$.
\end{theorem}
To establish this theorem, we decompose the score $n^{-1/2}\sum_{i=1}^n d_i^*\varepsilon_i$ into the sum of two
uncorrelated terms,
\[
n^{-1/2} \sum_{i=1}^{n} d_{i}^{\ast}\varepsilon_{i} = n^{-1/2}\sum_{i=1}^{n}
\left(  A_{i}- \mu_A\right)  ^{\top}e_{i} + n^{-1/2}
\sum_{i=1}^{n} \mu_A  ^{\top}e_{i},
\]
and observe that the first term on the right-hand side here forms a martingale
difference sequence while the second term is asymptotically normal by assumption. Combining these facts, we are able to obtain asymptotic normality of the sum; see the detailed proof in the Appendix.

Like Theorem \ref{L1}, this theorem establishes two key facts as well. To see both of them, observe that the term $\sigma_{d\varepsilon}^2$ appearing in the convergence result \eqref{linear-exp} can be more explicitly rewritten as
\begin{equation}
\sigma_{d\varepsilon}^2 = n^{-1}\sum_{i=1}^{n}\mathbb{E}\left[  (d_{i}^*)^2 \varepsilon_{i}^{2}\right]  + \mu_A^{\top}\left[
\operatorname{Var}\left( n^{-1/2} \sum_{i=1}^{n}e_{i}\right)  - n^{-1}\sum_{i=1}^{n}
\operatorname{Var}\left(  e_{i}\right)  \right]  \mu_A.
\label{Variance-sc}%
\end{equation}
This expression in turn implies that the asymptotic variance of the OLS estimator now depends on the correlation across the vectors $e_i$, which means that {\em heteroscedasticity} of the regression errors forces OLS variance estimators to be adjusted for {\em correlation} across the errors. On the other hand, the expression \eqref{Variance-sc} also demonstrates that it suffices to adjust the variance estimators only for correlation across the random variables $\mu_A^{\top}e_i$, instead of correlation across the full vectors $e_i$. The latter point might seem like a minor technicality but it in fact plays an interesting role in models with heterogeneous treatment effects. Indeed, when $d_i\in\{0,1\}$ represents the treatment assignment status, $(y_i(1),y_i(0))$ represents the pair of potential outcomes with and without treatment, so that $y_i = y_i(0) + d_i(y_i(1)-y_i(0))$, and $w_i$ consists of a non-zero constant only, it follows that $\varepsilon_i$ takes the form \eqref{eq:heteroex}, which matches the setting here with $e_{i} = (y_{i}(0)-\mathbb E[y_i(0)],y_{i}(1)-y_{i}(0) - \mathbb E[y_{i}(1)-y_{i}(0)])^{\top}$ and $\sigma(d_i)=(1,d_i)^{\top}$. Hence, $\mu_A = (0,\sigma_d^2)^{\top}$, and so the expression for the asymptotic variance of $\sqrt n(\hat\beta - \beta)$ reduces to the probability limit of
$$
\frac{1}{n\sigma_{d}^{4}}\sum_{i=1}^{n} \mathbb{E}[(d_{i} - \mu_d)^{2} \varepsilon_{i}^{2}] +
\text{Var}\left( n^{-1/2} \sum_{i=1}^{n} (y_{i}(1)-y_{i}(0))\right)  - n^{-1}\sum_{i=1}^{n}
\text{Var}\left(  y_{i}(1)-y_{i}(0)\right),
$$
as previewed in the previous section; see expression \eqref{eq: variance treatment effect formula} there. In turn, this expression means that it suffices to adjust OLS variance estimation for correlation across treatment effect, and there is no need to worry about correlation of potential outcomes within any given treatment arm. For example, whenever treatment effects $y_i(1)-y_i(0)$ are independent across $i$, it suffices to use the usual heteroscedasticity-robust variance formulas, even if regression errors $\varepsilon_i$ are correlated. Note, however, that it is {\em necessary} to use heteroscedasticity-robust variance formulas even if the treatment effects are i.i.d. (and not just independent), as the formula $s^{2}\left(\mathbf{X}^{\top}\mathbf{X}\right)^{-1}$ is not valid in this case.  Moreover, the same results apply even if $w_i$'s include non-constant controls as well, as the second component of $e_i$'s remains the same in this case.

Finally, we note that as long as the form of correlation across random variables $\mu_A^\top e_i$ is known, estimation of the asymptotic variance based on Theorem \ref{L3} is conceptually straightforward. For example, if treatment effects are clustered, it suffices to use \cite{moulton1986random}'s  or \cite{LiangZeger86}'s formulas assuming that regression errors are clustered at the same level (even though they could be clustered at a different level because of the clustering of potential outcomes without treatment, for example). For brevity of the paper, we do not provide formal statement of such results, as they are case-specific and depend on the form of the correlation structure, e.g. time series versus cluster versus spatial dependence.

%Unfortunately, the asymptotic distribution of the second
%term on the right hand side can be affected by dependence in the series
%$(e_{i})_{i\leq n}$ unless we know that $\mathbb{E}[A^{\ast}]=0$. Finally, we
%note that to recover the asymptotic variance formula for a heterogenous treatment
%effects model stated in Section \ref{sec:intuition}, we need only apply Lemma
%\ref{L3} with $e_{i} = (y_{i}(0),y_{i}(1)-y_{i}(0))^{\top}$ and $\sigma(d)
%\equiv(1,x)^{\top}$.

\section{Group-Level Randomization \label{sec:group}}

A key assumption behind the results of Section \ref{sec:main} is that the
regressor of interest is independent and identically distributed across units.
In randomized controlled trials, such an assumption is satisfied for
completely randomized assignments but can fail under other randomization
protocols. For instance, in randomized controlled trials the
i.i.d.\ assumption on the treatment fails when treatment is assigned at a
group level. Motivated by this challenge, we suppose in this section that
\begin{equation}
y_{i,j}=d_{j}\beta+w_{i,j}^{\top}\gamma+\varepsilon_{i,j}, \label{M2}%
\end{equation}
where the index $j=1,\ldots n_{g}$ denotes the group membership, the index
$i=1,\ldots,n_{j}$ denotes units within group $j$, $n_{g}$ is the number of
groups, $n_{j}$ is the number of units within group $j$, $d_j$ denotes the regressor of interest which is invariant within group $j$, $w_{i,j}$ denotes the vector of controls, and $\varepsilon_{i,j}$ is a mean-zero regression error. We continue to assume that the first component of each $w_{i,j}$ is a non-zero constant and also continue to employ $n$ as the total number of observations, $n = \sum_{j\leq n_{g}}
n_{j}$.

We emphasize at this point that the index $j$ distinguishes the level at which
the regressor $d_{j}$ varies, but has no special significance for other
variables. In particular, we do not insist that any (potential) clustering
structure of the errors $((\varepsilon_{i,j})_{i\leq n_{j}})_{j\leq n_{g}}$ be
the same as the group structure specified by the regressor $d_{j}$. In other
words, $((\varepsilon_{i,j})_{i\leq n_{j}})_{j\leq n_{g}}$ may have a
dependence structure completely different from group structure determined by
$j$ -- e.g., $\varepsilon_{i_{1},j_{1}}$ may be correlated with $\varepsilon
_{i_{2},j_{2}}$ when $j_{1} \neq j_{2}$. This discussion will be formalized in
the nature of the regularity conditions below, which do not include, for
example, the random effects type specification as in \cite{moulton1986random}.
Instead, as in the previous section, we will rely on a martingale structure
based on $(d_{j})_{j\leq n_{g}}$ that will be crucial for understanding the
asymptotic distribution of $\hat \beta$.
%\footnote{From the simple time series linear regression model (\ref{simple-regression}), we have noted that the $s^{2}\left(  \mathbf{X}^{\top}\mathbf{X}\right)  ^{-1}$ formula is equally valid as the $\left(  \mathbf{X}^{\top}\mathbf{X}\right) ^{-1}\mathbf{X}^{\top}\Omega \mathbf{X}\left(  \mathbf{X}^{\top}\mathbf{X}% \right)  ^{-1}$ formula. Likewise, we can see that \cite{moulton1986random}'s formula can be used in two different ways if the errors $((\varepsilon _{i,j})_{i\leq n_{j}})_{j\leq n_{g}}$ have a clustering structure different from the treatment variables $(x_{j})_{j\leq n_{g}}$ to construct valid standard error estimator. In other words, the formula using the group structure on the treatment variable $x_{j}$ and the one using the clustering structure on the errors are both valid. We suspect that in the RCT, it may be more pragmatic to use the group structure in the treatment variable, because the clustering structure of the error term is often unknown to the practitioners.}

Following our discussion in the previous section, we consider cases of strong exogeneity and conditional heteroscedasticity separately. We first consider the case of strong exogeneity. In order to study the asymptotic properties of $\hat \beta$, we first need to
revise Assumptions \ref{A1} and \ref{A3} to account for the group structure
and a possible within-group correlation. To this end, we let $\kappa_{n}%
\equiv(\sum_{j\leq n_{g}}n_{j}^{2})^{1/2}$ and $S_{\varepsilon}^2\equiv
n^{-1}\sum_{j\leq n_{g}}\mathbb E[(\sum_{i\leq n_{j}}\varepsilon
_{i,j})^2]$ and impose the following assumptions.

\begin{assumption}
\label{A1'}(i)\ The random variables $d_{j}$, $1\leq j\leq n_g$, are i.i.d. with mean $\mu_d$ and variance $\sigma_d^2$; (ii) $(d_{j})_{j\leq
n_g}$ is independent of $((w_{i,j}^{\top})_{i\leq n_j})_{j\leq n_g};$  (iii)\ $\max_{j\leq n_g}\mathbb{E}[|d_{j}|^{4}]\leq K$; (iv) $\sigma_{d}^{2}\geq K^{-1}$; (v) $(d_{j})_{i\leq
n_g}$ is independent of $((\varepsilon_{i,j})_{i\leq n_j})_{j\leq n_g}$.
\end{assumption}

\begin{assumption}
\label{A3'}(i)\ $n^{-1}\sum_{j=1}^{n_{g}}\sum_{i=1}^{n_{j}}w_{i,j}\varepsilon_{i,j}=o_{p}(S_{\varepsilon}n^{1/2}/\kappa_n)$; (ii) $S_{\varepsilon}^{-2}n^{-1}\sum_{j=1}^{n_g}(\sum_{i=1}^{n_j}\varepsilon_{i,j})^2\to_p 1$; (iii) there exists a constant $\delta_4\in(0,2]$ such that $n^{-1-\delta_4/2}\sum_{j=1}^{n_g}(|\sum_{i=1}^{n_j}\varepsilon_{i,j}|)^{2+\delta_4}=o_p(S_{\varepsilon}^{2+\delta_4})$; (iv) $\kappa_n/n=o(1)$.
\end{assumption}

Assumption \ref{A1'} requires mild moment restrictions and that the regressor
of interest $d_{j}$ be strongly exogenous in the sense that it be
independent of the errors and other regressors. To make sense of Assumption \ref{A3'}, assume that each group $j$ has the same size $n_j$ that is independent of $n$. In this case, $\kappa_n$ is of order $\sqrt n$ and $S_{\varepsilon}^2$ is typically of order one. In turn, the latter implies that Assumption \ref{A3'}(i) reduces to $n^{-1}\sum_{j\leq n_{g}}\sum_{i\leq n_{j}}w_{i,j}\varepsilon_{i,j}=o_{p}(1)$, which is similar to Assumption \ref{A3}(i), and Assumption \ref{A3'}(iii) reduces to $n^{-1-\delta_4/2}\sum_{j=1}^{n_g}\sum_{i=1}^{n_j}|\varepsilon_{i,j}|^{2+\delta_4} = o_p(1)$, which is satisfied as long as $\max_{1\leq j\leq n_g}\max_{1\leq i\leq n_j}|\varepsilon_{i,j}|^{2+\delta_4} \leq K$. In addition, Assumption \ref{A3'}(iv) reduces $n^{-1/2}=o(1)$ and is satisfied automatically and Assumption \ref{A3'}(ii) can be regarded as a law of large numbers. Thus, Assumption \ref{A3'} in general requires that the size of groups does not increase too fast. Note also that, as previously claimed, this assumption does not require $((w_{i,j}^{\top},\varepsilon
_{i,j})_{i\leq n_{j}})_{j\leq n_{g}}$ to be independent across $j$.

We are now ready to derive the asymptotic distribution of the OLS estimator $\hat\beta$ in the strongly exogenous case with a group-level assignment. In the statement of the result, we impose
Assumption \ref{A2}, which should be interpreted to hold with $\sum
_{j=1}^{n_{g}}\sum_{i=1}^{n_{j}}$ in place of $\sum_{i=1}^{n}$ in Assumption
\ref{A2}(i) and $\max_{j\leq n_{g}}\max_{i\leq n_{j}}$ in place of
$\max_{i\leq n}$ in Assumption \ref{A2}(ii). Also, we denote $d_j^* \equiv d_j - \mu_d$ for all $1\leq j\leq n_g$.

\begin{theorem}
\label{L2}Let Assumptions \ref{A2}, \ref{A1'} and \ref{A3'} hold. Then
\begin{equation}
\frac{\sqrt{n}(\hat{\beta}-\beta)}{S_{\varepsilon}/\sigma_{d}}%
=\frac{n^{-1/2}\sum_{j=1}^{n_{g}}d_{j}^*\sum_{i=1}^{n_{j}}\varepsilon_{i,j}}{\sigma_{d}S_{\varepsilon}}+o_{p}(1)\to_d N(0,1).
\label{P_L2-1}%
\end{equation}
\end{theorem}
The implications of this theorem are following. First, the OLS estimator $\hat \beta$ is asymptotically normally
distributed under mild moment conditions and under fairly general assumptions
on the dependence structure of $((w_{i,j}^{\top},\varepsilon_{i,j})_{i\leq
n_{j}})_{j\leq n_{g}}$. Second, the asymptotic variance of $\hat \beta$ is
given by $S_{\varepsilon}^2/\sigma_{d}^{2}$. In particular, since
$S_{\varepsilon}^2 = n^{-1}\sum_{j\leq n_{g}} \text{Var}(\sum_{i\leq n_{j}%
}\varepsilon_{i,j})$, when computing standard errors for $\hat \beta$ we need
only account for possibly within-group $j$ correlation \emph{even if}
$((\varepsilon_{i,j})_{i\leq n_{j}})_{j\leq n_{g}}$ is dependent across $j$.
Importantly, the group structure is determined solely by the variables
$(d_{j})_{j\leq n_{g}}$ and hence is known, considerably simplifying
estimation. For instance, in a randomized controlled trial with constant
effects, Theorem \ref{L2} implies we may, for example, employ
\cite{moulton1986random}'s or \cite{LiangZeger86} formulas clustered at the level at which treatment
was assigned.
%Andres: I am commenting the HAC thing here out, because it is not clear whether this works if the size of the cluster is fixed. (\sum_{i\leq n_{j}}\varepsilon_{i,j})$ by a HAC estimator. In panel models with random effects, this key quantity can be estimated using \cite{moulton1986random}'s formula.
We again
emphasize, however, that similar conclusions do not apply for $\hat \gamma$
whose standard errors and asymptotic normality may depend on the dependence
structure of $((w_{i,j}^{\top},\varepsilon_{i,j})_{i\leq n_{j}})_{j\leq n_{g}%
}$ across $j$.

Next, we consider the case of conditional heteroscedasticity. Denote $S_{e,1}^2\equiv n^{-1}\sum_{j=1}^{n_g}\mathbb E[((A_j - \mu_A)^{\top}\sum_{i=1}^{n_j}e_{i,j})^2]$ and $S_{e,2}^2\equiv \mathbb E[(n^{-1/2}\sum_{i=1}^n\mu_A^{\top}e_i)^2]$. Note that $S_{e,2}$ here actually coincides with $\sigma_{e,2}$ in the previous section. Within this context, we impose the following assumptions.

\begin{assumption}
\label{A8}(i) $(d_{j})_{j\leq n_g}$ is independent of $((e_{i,j}^\top)_{i\leq n_j})_{j\leq n_g}$; (ii) the functions $\sigma_l$, $1\leq l\leq L$, are bounded; (iii) $S_{e,1} / S_{\varepsilon} \geq K^{-1}$.
\end{assumption}
\begin{assumption}
\label{A9} (i) $\|n^{-1}\sum_{j=1}^{n_g} (\sum_{i=1}^{n_j}e_{i,j})(\sum_{i=1}^{n_j} e_{i,j})^{\top} - n^{-1}\sum_{j=1}^{n_g}\mathbb E[(\sum_{i=1}^{n_j}e_{i,j})(\sum_{i=1}^{n_j}e_{i,j})^{\top}]\| = o_p(S_{e,1}^2)$; (ii) $n^{-2}\sum_{j=1}^{n_g}\|\sum_{i=1}^{n_j} e_{i,j}\|^{4} = o_p(S_{e,1}^4)$; (iii) $S_{e,2}^{-1}n^{-1/2}\sum_{i=1}^n\mu_A^{\top}e_i \to_d N(0,1)$.
\end{assumption}

These assumptions naturally extend Assumptions \ref{A4} and \ref{A5new} in the previous section to allow for group-level assignments.

The next theorem derives the asymptotic distribution of the OLS estimator $\hat\beta$ in the case of conditional heteroscedasticity with a group-level assignment.

\begin{theorem}\label{thm: clustering plus hetero}
Let Assumptions \ref{A2}, \ref{A1'}(i)-(iv), \ref{A3'}, \ref{A8}, and \ref{A9} hold. Then
\begin{equation}\label{eq: clustering plus hetero convergence}
\frac{\sqrt{n}(\hat{\beta}-\beta)}{S_{d \varepsilon}/\sigma_{d}^2}%
=\frac{n^{-1/2}\sum_{j=1}^{n_{g}}d_j^*\sum_{i=1}^{n_{j}}\varepsilon_{i,j}}{S_{d \varepsilon}}+o_{p}(1)\to_d N(0,1),
\end{equation}
where $S_{d\varepsilon} \equiv S_{e,1}^2 + S_{e,2}^2$.
\end{theorem}
This theorem relates to Theorem \ref{L2} in the same way as Theorem \ref{L3} relates to Theorem \ref{L1}. In particular, noting that the term $S_{d\varepsilon}^2$ appearing in this theorem can be more explicitly rewritten as
$$
S_{d\varepsilon}^2 \equiv n^{-1}\sum_{j=1}^{n_g}\mathbb E\left[ \left(d_j^*\sum_{i=1}^{n_j} \varepsilon_{i,j}\right)^2 \right] + \mu_A^\top\left[ \operatorname{Var}\left(n^{-1/2}\sum_{j=1}^{n_g}\sum_{i=1}^{n_j}e_{i,j}\right) - n^{-1}\sum_{j=1}^{n_g}\operatorname{Var}\left(\sum_{i=1}^{n_j} e_{i,j}\right) \right] \mu_A,
$$
we conclude that because of conditional heteroscedasticity, variance estimators that are clustered at the group level at which the regressor $d_j$ is assigned may not be valid if the errors $e_{i,j}$ are correlated across groups $j$. On the other hand, in the context of estimation with heterogeneous treatment effects, such estimators are valid if the treatment effects are uncorrelated across these groups.

%\section{Conclusion\label{sec:conclusion}}

%We studied the asymptotic properties of the least squares estimator for the
%coefficient of a strongly exogenous regressor, such as treatment in a
%randomized controlled trial. In our baseline case, we showed that textbook
%homoscedastic standard errors are valid under general dependence structures in
%the errors and other regressors. A similar conclusion also holds when the
%regressor has a group structure, which has implications for clustered standard
%errors. Underlying our analysis is a common martingale structure that
%delivered our main results. Since in randomized controlled trials
%practitioners often know the distribution of treatment assignment, we expect
%similar martingale arguments to be potentially applicable to more complex
%randomization protocols.

\bigskip

{\small
\bibliographystyle{econometrica}
\bibliography{RCT}
}

\newpage

\appendix

\begin{center}
{\huge Appendix}
\end{center}

\section{Proof of the main results\label{APP_1}}

\noindent \textsc{Proof of Theorem \ref{L1}}. Define $\mathbf{\tilde{D}\equiv
D}-\mathbf{1}_{n\times1}(n^{-1}\mathbf{D}^{\top}\mathbf{1}_{n\times1})$ and
$\mathbf{M}_{W}\equiv \mathbf{I}_{n}-\mathbf{W}(\mathbf{W}^{\top}%
\mathbf{W})^{-1}\mathbf{W}^{\top}$. Then, given that the matrix $\mathbf W$ includes a non-zero constant column by Assumption \ref{A2}(iii), it follows that $\mathbf{1}_{n\times1}^{\top}\mathbf{M}_{W}=0$. Therefore, by the Frisch-Waugh-Lovell theorem,
\begin{equation}
\hat{\beta}-\beta=(\mathbf{D}^{\top}\mathbf{M}_{W}\mathbf{D})^{-1}%
(\mathbf{D}^{\top}\mathbf{M}_{W}%
\bm{\epsilon}%
)=(\mathbf{\tilde{D}}^{\top}\mathbf{M}_{W}\mathbf{\tilde{D}})^{-1}%
(\mathbf{\tilde{D}}^{\top}\mathbf{M}_{W}%
\bm{\epsilon}%
). \label{P_L1_1}%
\end{equation}
Also, denoting $\overline d_n\equiv n^{-1}\sum_{i=1}^n d_i$, we have
$$
\mathbb E[|\overline d_n - \mu_d|^2] = \sigma_d^2/n \leq K/n = o(1)
$$
by Assumptions \ref{A1}(i,iii), and so $\overline d_n - \mu_d = o_p(1)$ by Markov's inequality. Hence,
$$
\overline d_n^2 - \mu_d^2 = (\overline d_n-\mu_d)(\overline d_n + \mu_d) = o_p(1)
$$
by Assumption \ref{A1}(iii) again. In addition,
$$
\mathbb E\left[\left| n^{-1}\sum_{i=1}^n(d_i^2 - \mathbb E[d_i^2]) \right|^{1+\delta_1/2}\right] \leq 2\sum_{i=1}^n\mathbb E\left[|d_i^2 - \mathbb E[d_i^2]|^{1+\delta_1/2}\right] / n^{1+\delta_1/2} = o(1)
$$
by the von Bahr-Esseen Inequality  (see Section 35.1.5 in \cite{Dasgupta2008}) and Assumptions \ref{A1}(i,iii). Hence, by Markov's inequality,
$$
n^{-1}\sum_{i=1}^n d_i^2 = n^{-1}\sum_{i=1}^n \mathbb E[d_i^2] + o_p(1),
$$
and so
\begin{equation}
n^{-1}\mathbf{\tilde{D}}^{\top}\mathbf{\tilde{D}} = n^{-1}\sum_{i=1}^n \mathbb E[d_i^2]  - \mu_d^2 + o_p(1)  =\sigma_{d}^{2}+o_{p}(1).
\label{P_L1_2}%
\end{equation}
Further,
\begin{equation}
n^{-1}\mathbf{\tilde{D}}^{\top}\mathbf{W}=n^{-1}\sum_{i=1}^{n}d_{i}^{\ast
}w_{i}-\left(  n^{-1}\sum_{i=1}^{n}d_{i}^{\ast}\right)  \left(  n^{-1}%
\sum_{i=1}^{n}w_{i}\right). \label{P_L1_3}%
\end{equation}
By Assumptions \ref{A1}(i, ii) and \ref{A2}(iii), we further obtain
\begin{align*}
\mathbb{E}\left[  \left \Vert n^{-1/2}\sum_{i=1}^{n}d_{i}^{\ast}w_{i}%
\right \Vert ^{2}\right]   &  =n^{-1}\sum_{i=1}^{n}\mathbb{E}\left[
||d_{i}^{\ast}w_{i}||^{2}\right]  \leq n^{-1}\sum
_{i=1}^{n}\mathbb{E}[d_{i}^{2}]\mathbb{E}\left[  ||w_{i}||^{2}\right]  \leq K.
\end{align*}
Hence, by Markov's inequality,
\begin{equation}
n^{-1}\sum_{i=1}^{n}d_{i}^{\ast}w_{i}=O_{p}(n^{-1/2}). \label{P_L1_4}%
\end{equation}
Similarly, we can use Assumptions \ref{A1}(i, iii) and \ref{A2}(ii) to show%
\begin{equation}
\left(  n^{-1}\sum_{i=1}^{n}d_{i}^{\ast}\right)  \left(  n^{-1}\sum_{i=1}%
^{n}w_{i}\right)  =O_{p}(n^{-1/2}), \label{P_L1_5}%
\end{equation}
which together with (\ref{P_L1_3}) and (\ref{P_L1_4}) implies that%
\begin{equation}
n^{-1}\mathbf{\tilde{D}}^{\top}\mathbf{W}=O_{p}(n^{-1/2}). \label{P_L1_6}%
\end{equation}
Combining this result with \eqref{P_L1_2} and using Assumptions \ref{A2}(i), we then have
\begin{equation}
n^{-1}\mathbf{\tilde{D}}^{\top}\mathbf{M}_{W}\mathbf{\tilde{D}}=n^{-1}%
\mathbf{\tilde{D}}^{\top}\mathbf{\tilde{D}}-n^{-1}\mathbf{\tilde{D}}^{\top
}\mathbf{W}(n^{-1}\mathbf{W}^{\top}\mathbf{W})^{-1}n^{-1}\mathbf{W}^{\top
}\mathbf{\tilde{D}}=\sigma_{d}^{2}+o_{p}(1). \label{P_L1_7}%
\end{equation}
For the term $\mathbf{\tilde{D}}^{\top}\mathbf{M}_{W}%
%TCIMACRO{\TeXButton{epsilon}{\bm{\epsilon}}}%
%BeginExpansion
\bm{\epsilon}%
%EndExpansion
$ in the numerator of $\hat{\beta}-\beta$ in \eqref{P_L1_1}, we have%
\begin{align}
n^{-1/2}\mathbf{\tilde{D}}^{\top}\mathbf{M}_{W}%
%TCIMACRO{\TeXButton{epsilon}{\bm{\epsilon}}}%
%BeginExpansion
\bm{\epsilon}%
%EndExpansion
&  =n^{-1/2}\mathbf{\tilde{D}}^{\top}%
\bm{\epsilon}%
-n^{-1/2}\mathbf{\tilde{D}}^{\top}\mathbf{W}(\mathbf{W}^{\top}\mathbf{W}%
)^{-1}\mathbf{W}^{\top}%
\bm{\epsilon}%
%EndExpansion
\nonumber \\
&  =n^{-1/2}\sum_{i=1}^{n}d_{i}^{\ast}\varepsilon_{i}-\left(  n^{-1/2}%
\sum_{i=1}^{n}d_{i}^{\ast}\right)  \left(  n^{-1}\sum_{i=1}^{n}\varepsilon
_{i}\right)  +o_{p}(1)\nonumber \\
&  =n^{-1/2}\sum_{i=1}^{n}d_{i}^{\ast}\varepsilon_{i}+o_{p}(1), \label{P_L1_8}%
\end{align}
where the second equality follows by the definition of $\mathbf{\tilde{D}}$,
(\ref{P_L1_6}), and Assumptions \ref{A2}(i) and \ref{A3}(i), and the third
equality follows by Assumptions \ref{A1}(i, iii) and \ref{A3}(iii) and Markov's inequality. Therefore,
\begin{equation}\label{eq: I did it}
\sqrt n(\hat\beta - \beta) = \frac{n^{-1/2}\sum_{i=1}^n d_i^*\varepsilon_i}{\sigma_d^2} + o_p(1)
\end{equation}
by Assumption \ref{A1}(iv), and so
\begin{equation}\label{eq: I did it again}
\frac{\sqrt n(\hat\beta - \beta)}{\sigma_{\varepsilon}/\sigma_d} = \frac{n^{-1/2}\sum_{i=1}^n d_i^*\varepsilon_i}{\sigma_d\sigma_{\varepsilon}} + o_p(1)
\end{equation}
by Assumptions \ref{A1}(iii) and \ref{A3}(iv).

We next derive the asymptotic distribution of $n^{-1/2}\sum_{i=1}^{n}%
d_{i}^{\ast}\varepsilon_{i}/(\sigma_d\sigma_{\varepsilon})$. Let $\mathcal{F}_{i,n}$ denote the filtration
generated by $(%
%TCIMACRO{\TeXButton{epsilon}{\bm{\epsilon}}}%
%BeginExpansion
\bm{\epsilon}%
%EndExpansion
^{\top},((d_{j})_{j\leq i})^{\top})$. Then by Assumptions \ref{A1}(iii,iv) and
\ref{A3}(iii, iv), $n^{-1/2}d_{i}^{\ast}\varepsilon_{i}/(\sigma_{d}\sigma
_{\varepsilon})$ has finite second moment and
\begin{equation}
\mathbb{E}\left[  \frac{n^{-1/2}d_{i}^{\ast}\varepsilon_{i}}{\sigma_{d}%
\sigma_{\varepsilon}}|\mathcal{F}_{i-1,n}\right]  =\frac{n^{-1/2}%
\varepsilon_{i}}{\sigma_{d}\sigma_{\varepsilon}}\mathbb{E}\left[  \left.
d_{i}^{\ast}\right \vert \mathcal{F}_{i-1,n}\right]  =\frac{n^{-1/2}%
\varepsilon_{i}}{\sigma_{d} \sigma_{\varepsilon}}\mathbb{E}\left[  d_{i}%
^{\ast}\right]  =0 \label{P_L1_9a}%
\end{equation}
almost surely, which implies that $n^{-1/2}d_{i}^{\ast}\varepsilon_{i}%
/(\sigma_{d}\sigma_{\varepsilon})$ is a martingale difference\ array with
respect to $\mathcal{F}_{i,n}$.\ Next, observe that Assumptions \ref{A1}(i,v)
and \ref{A3}(ii,iv) yield
\begin{equation}
\sum_{i=1}^{n}\mathbb{E}\left[  \left(  \frac{n^{-1/2}d_{i}^{\ast}%
\varepsilon_{i}}{\sigma_{d}\sigma_{\varepsilon}}\right)  ^{2}|\mathcal{F}%
_{i-1,n}\right]  =\sigma_{\varepsilon}^{-2}n^{-1}\sum_{i=1}^{n}\varepsilon
_{i}^{2}\rightarrow_{p}1. \label{P_L1_9}%
\end{equation}
Moreover for any $\eta>0$, Assumptions \ref{A1}(iii,iv,v) and \ref{A3}(iii,iv) allow us to
conclude that for $\delta \equiv \min(\delta_1,\delta_2)$,
\begin{align}
&  \sum_{i=1}^{n}\mathbb{E}\left[  \allowbreak \left.  \left(  \frac
{n^{-1/2}d_{i}^{\ast}\varepsilon_{i}}{\sigma_{d}\sigma_{\varepsilon}}\right)
^{2}1\left \{  \left \vert \frac{n^{-1/2}d_{i}^{\ast}\varepsilon_{i}}{\sigma
_{d}\sigma_{\varepsilon}}\right \vert >\eta \right \}  \right \vert \mathcal{F}%
_{i-1,n}\right] \nonumber \\
&  \leq \frac{1}{\eta^{\delta}}\sum_{i=1}^{n}\mathbb{E}\left[  \left.
\left \vert \frac{n^{-1/2}d_{i}^{\ast}\varepsilon_{i}}{\sigma_{d}%
\sigma_{\varepsilon}}\right \vert^{2+\delta}\right \vert
\mathcal{F}_{i-1,n}\right] \nonumber \\
&  =\frac{1  }{\eta^{\delta
}(\sigma_{d}\sigma_{\varepsilon})^{2+\delta}n^{1+\delta/2}}\sum_{i=1}%
^{n}\mathbb{E}\left[  |d_{i}^{\ast}|^{2+\delta}\right]|\varepsilon_{i}\allowbreak|^{2+\delta}\leq \frac{K}{\eta^{\delta}%
n^{\delta/2}}n^{-1}\sum_{i=1}^{n}|\varepsilon_{i}\allowbreak|^{2+\delta} = o_p(1).
\label{P_L1_10}%
\end{align}
In view of (\ref{P_L1_9}) and (\ref{P_L1_10}), we can invoke the martingale
central limit theorem (see, e.g., Corollary 3.1 in \cite{HallHeyde1980}) to
conclude that%
\begin{equation}
\frac{n^{-1/2}\sum_{i=1}^{n}d_{i}^{\ast}\varepsilon_{i}}{\sigma_{d}%
\sigma_{\varepsilon}}\rightarrow_{d}N(0,1). \label{P_L1_12}%
\end{equation}
The claim of the theorem follows from combining this result with \eqref{eq: I did it again}.\hfill$Q.E.D.$

\bigskip

\noindent \textsc{Proof of Corollary \ref{T1}}.
We first proof that $\lambda_{\min}(n^{-1}\mathbf{X}^{\top}\mathbf{X})\geq K^{-1} + o_p(1)$. To do so, observe that
$$
n^{-1}\mathbf{D}^{\top}\mathbf{D} = \mu_d^2 + \sigma_d^2 + o_p(1)
$$
by Assumptions \ref{A1}(i,iii) and the law of large numbers. Also, denoting $d_i^* = d_i -\mu_d$ for all $1\leq i\leq n$, we have
$$
n^{-1}\mathbf{D}^{\top}\mathbf{W} = n^{-1}\sum_{i=1}^nd_iw_i = \mu_d n^{-1}\sum_{i=1}^n w_i + n^{-1}\sum_{i=1}^n d_i^*w_i = \mu_dn^{-1}\sum_{i=1}^nw_i + o_p(1),
$$
where the last equality follows from \eqref{P_L1_4} in the proof of Theorem \ref{L1}. Hence,
$$
n^{-1}\mathbf{X}^{\top}\mathbf{X} = n^{-1}(\mathbf{D},\mathbf{W})^{\top}(\mathbf{D},\mathbf{W}) = n^{-1}\sum_{i=1}^n(\mu_d,w_i^{\top})^{\top}(\mu_d,w_i^{\top}) + (\sigma_d,\mathbf{0}_{d_w\times 1}^{\top})^{\top}(\sigma_d,\mathbf{0}_{d_w\times 1}^{\top}) + o_p(1),
$$
where $d_w$ is the dimension of the vectors $w_i$. Now, denote
$$
R_n\equiv \min\left(\frac{\sqrt{3\lambda_{\min}(n^{-1}\sum_{i=1}^nw_iw_i^{\top})}}{8|\mu_d|},\frac{1}{2}\right)
$$
and fix any $a_1\in\mathbb R$ and $a_2\in\mathbb R^{d_w}$ such that $a_1^2 + \|a_2\|^2 = 1$. If $|a_1| > R_n$, then
$$
(a_1,a_2^{\top})\left( n^{-1}\sum_{i=1}^n(\mu_d,w_i^{\top})^{\top}(\mu_d,w_i^{\top}) + (\sigma_d,\mathbf{0}_{d_w\times 1}^{\top})^{\top}(\sigma_d,\mathbf{0}_{d_w\times 1}^{\top}) \right)(a_1,a_2^{\top})^{\top} \geq a_1^2\sigma_d^2 \geq R_n^2\sigma_d^2 \geq K^{-1}
$$
by Assumptions \ref{A1}(iii,iv) and \ref{A2}(i). If, on the other hand, $|a_1| \leq R_n$, then $\|a_2\|^2 = 1 - a_1^2 \geq 1 - 1/4 = 3/4$, and so
$$
|a_1| \leq \frac{\|a_2\|\sqrt{\lambda_{\min}(n^{-1}\sum_{i=1}^nw_iw_i^{\top})}}{4|\mu_d|}.
$$
The latter in turn implies via Jensen's inequality that
\begin{align*}
& (a_1,a_2^{\top})\left( n^{-1}\sum_{i=1}^n(\mu_d,w_i^{\top})^{\top}(\mu_d,w_i^{\top}) + (\sigma_d,\mathbf{0}_{d_w\times 1}^{\top})^{\top}(\sigma_d,\mathbf{0}_{d_w\times 1}^{\top}) \right)(a_1,a_2^{\top})^{\top}  \\
& \qquad \geq n^{-1}\sum_{i=1}^n (a_1\mu_d + a_2^{\top}w_i)^2 \geq n^{-1}\sum_{i=1}^n (a_2^{\top}w_i)^2 - 2|a_1\mu_d| n^{-1}\sum_{i=1}^n|a_2^{\top}w_i| \\
& \qquad \geq n^{-1}\sum_{i=1}^n (a_2^{\top}w_i)^2 - 2|a_1\mu_d| \sqrt{n^{-1}\sum_{i=1}^n(a_2^{\top}w_i)^2} \\
& \qquad = \sqrt{n^{-1}\sum_{i=1}^n(a_2^{\top}w_i)^2}\left(\sqrt{n^{-1}\sum_{i=1}^n(a_2^{\top}w_i)^2} - 2|a_1\mu_d|\right) \\
&\qquad \geq  \|a_2\|\sqrt{\lambda_{\min}\left(n^{-1}\sum_{i=1}^nw_iw_i^{\top}\right)}\left(\|a_2\|\sqrt{\lambda_{\min}\left(n^{-1}\sum_{i=1}^nw_iw_i^{\top}\right)} - 2|a_1\mu_d|\right) \\
& \qquad \geq 2^{-1} \|a_2\|^2\lambda_{\min}\left(n^{-1}\sum_{i=1}^nw_iw_i^{\top}\right) \geq (3/8)\lambda_{\min}\left(n^{-1}\sum_{i=1}^nw_iw_i^{\top}\right) \geq K^{-1}
\end{align*}
by Assumption \ref{A2}(i). Hence, it follows that
$$
\lambda_{\min}\left(n^{-1}\sum_{i=1}^n(\mu_d,w_i^{\top})^{\top}(\mu_d,w_i^{\top}) + (\sigma_d,\mathbf{0}_{d_w\times 1}^{\top})^{\top}(\sigma_d,\mathbf{0}_{d_w\times 1}^{\top})\right)\geq K^{-1},
$$
and so
\begin{equation}\label{eq: min eig value}
\lambda_{\min}(n^{-1}\mathbf{X}^{\top}\mathbf{X}) \geq K^{-1} + o_p(1),
\end{equation}
as required.

Next, we prove that
$s^{2}$ is consistent for
$\sigma_{\varepsilon}^{2}$. To this end, note that by Assumptions \ref{A1}(i, iii, v) and \ref{A3}(iii),
\begin{equation}
\mathbb{E}\left[  \left(  n^{-1}\sum_{i=1}^{n}d_{i}^{\ast}\varepsilon
_{i}\right)  ^{2}\right]  =n^{-2}\sum_{i=1}^{n}\mathbb{E}\left[  d_{i}^{\ast
2}\varepsilon_{i}^{2}\right]  \leq Kn^{-1}. \label{P_T1_5}%
\end{equation}
Hence, by Assumptions \ref{A2}(iii) and \ref{A3}(i) and Markov's
inequality
\[
n^{-1}\mathbf{D}^{\top}%
%TCIMACRO{\TeXButton{epsilon}{\bm{\epsilon}}}%
%BeginExpansion
\bm{\epsilon}%
%EndExpansion
=n^{-1}\sum_{i=1}^{n}d_{i}^{\ast}\varepsilon_{i}+\mathbb{E}[d_{i}]n^{-1}%
\sum_{i}\varepsilon_{i}=o_{p}(1),
\]
which combined with Assumption \ref{A3}(i) further implies that%
\begin{equation}
n^{-1}\mathbf{X}^{\top}\bm{\epsilon} = n^{-1}(\mathbf{D},\mathbf{W})^{\top}%
%TCIMACRO{\TeXButton{epsilon}{\bm{\epsilon}}}%
%BeginExpansion
\bm{\epsilon}%
%EndExpansion
=o_{p}(1). \label{P_T1_6}%
\end{equation}
Combining this bound with \eqref{eq: min eig value} gives
\begin{equation}
\hat{\theta} - \theta = (n^{-1}\mathbf{X}^{\top}\mathbf{X})^{-1}(n^{-1}\mathbf{X}^{\top}
\bm{\epsilon}%
%EndExpansion
\mathbf{)}=o_{p}(1). \label{P_T1_7}%
\end{equation}
In addition, $\mathbf{Y}-\mathbf X\hat{\theta}=%
%TCIMACRO{\TeXButton{epsilon}{\bm{\epsilon}}}%
%BeginExpansion
\bm{\epsilon}%
%EndExpansion
-\mathbf X(\hat{\theta}-\theta)$. Therefore,
\begin{align}
s^{2}  &  =n^{-1}(\mathbf{Y}-\mathbf X\hat{\theta})^{\top
}(\mathbf{Y}-\mathbf X\hat{\theta})  =n^{-1}(\bm{\epsilon}%
-\mathbf X(\hat{\theta}-\theta))^{\top}(\bm{\epsilon}
-\mathbf X(\hat{\theta}-\theta))\nonumber \\
& =n^{-1}%
\bm{\epsilon}
^{\top}%
\bm{\epsilon}%
-2n^{-1}(\hat{\theta}-\theta)^{\top}\mathbf X^{\top}%
\bm{\epsilon}%
%EndExpansion
+n^{-1}(\hat{\theta}-\theta)^{\top}\mathbf X^{\top}%
\mathbf X(\hat{\theta}-\theta)\nonumber \\
&  =\sigma_{\varepsilon}^{2}+o_{p}(1), \label{P_T1_8}%
\end{align}
where the last equality follows from (\ref{P_T1_6}) and (\ref{P_T1_7}) and Assumptions \ref{A1}(iii), \ref{A2}(ii) and \ref{A3}(ii). This finishes the proof of consistency of
$s^{2}$.

Now, defining $\mathbf{\tilde{D}\equiv
D}-\mathbf{1}_{n\times1}(n^{-1}\mathbf{D}^{\top}\mathbf{1}_{n\times1})$ to match the proof of Theorem \ref{T1} and recalling that $\mathbf{\breve{D}}=\mathbf{M}_{W}\mathbf D$, we have
\begin{align*}
n^{-1}\mathbf{\breve{D}}^{\top}\mathbf{\breve{D}}  &  =n^{-1}\mathbf{\tilde
{D}}^{\top}\mathbf{M}_{W}\mathbf{\tilde{D}}=n^{-1}\mathbf{\tilde{D}}^{\top
}\mathbf{\tilde{D}}-n^{-1}(\mathbf{\tilde{D}}^{\top}\mathbf{W})(\mathbf{W}%
^{\top}\mathbf{W})^{-1}(\mathbf{W}^{\top}\mathbf{\tilde{D}})\\
&  =n^{-1}\mathbf{\tilde{D}}^{\top}\mathbf{\tilde{D}}+o_{p}(1)=\sigma_{d}%
^{2}+o_{p}(1),
\end{align*}
where the third equality follows from \eqref{P_L1_6} and Assumption \ref{A2}(i), and
the fourth from \eqref{P_L1_2}. Together with \eqref{P_T1_8}, this gives the
first convergence result in \eqref{eq: asy variance estimator consistency} since $\sigma_{d}^{2}\geq K^{-1}$
by Assumption \ref{A1}(iii). The second convergence result follows from the first one, Theorem \ref{L1}, and Slutsky's
theorem.\hfill$Q.E.D.$

\bigskip

\noindent \textsc{Proof of Theorem \ref{L3}}.\  As in the proof of Theorem \ref{L1}, we have
\begin{equation}\label{eq: proof of thm 2 beginning}
\sqrt n(\hat\beta - \beta) = \frac{n^{-1/2}\sum_{i=1}^n d_i^*\varepsilon_i}{\sigma_d^2} + o_p(1);
\end{equation}
see Equation \eqref{eq: I did it} there and note that the derivation of \eqref{eq: I did it} did not rely on Assumption \ref{A1}(v), which we are not imposing here. Since $\sigma_{d\varepsilon}^2 \geq \sigma_{e,1}^2 \geq K^{-1}$ by Assumption \ref{A4}(iii) and $\sigma_d^2 \leq K$ by Assumption \ref{A1}(iii), \eqref{eq: proof of thm 2 beginning} implies that
$$
\frac{\sqrt n(\hat\beta - \beta)}{\sigma_{d\varepsilon}/\sigma_d^2} = \frac{n^{-1/2}\sum_{i=1}^n d_i^*\varepsilon_i}{\sigma_{d\varepsilon}} + o_p(1),
$$
which yields the equality in \eqref{linear-exp}.

We next derive the convergence result in \eqref{linear-exp}, i.e. we show that $\sigma_{d\varepsilon}^{-1}n^{-1/2}\sum_{i=1}^nd_i^*\varepsilon_i\to_d N(0,1)$. To do so, we write
$$
n^{-1/2}\sum_{i=1}^n d_i^*\varepsilon_i = n^{-1/2}\sum_{i=1}^n A_i^{\top}e_i = n^{-1/2}\sum_{i=1}^n (A_i - \mu_A)^{\top}e_i + n^{-1/2}\sum_{i=1}^n \mu_A^{\top}e_i
$$
and denote the first and the second terms on the right-hand side by $M_1$ and $M_2$, respectively. Also, we denote $\tilde A_i\equiv A_i - \mu_A$ for all $1\leq i\leq n$. In addition, denote $\hat\sigma_{e,1}^2\equiv n^{-1}\sum_{i=1}^n \mathbb E[(\tilde A_i^\top e_i)^2\mid \mathcal F_e]$, where $\mathcal F_e$ is the filtration generated by $(e_i^{\top})_{i\leq n}$. Then
\begin{align}
\hat\sigma_{e,1}^2
&=n^{-1}\sum_{i=1}^n e_i^{\top}\mathbb E[\tilde A_i\tilde A_i^\top]e_i
   =n^{-1}\sum_{i=1}^n \text{tr}(\mathbb E[\tilde A_i\tilde A_i^\top]e_ie_i^\top)
   =\text{tr}\left(n^{-1}\sum_{i=1}^n \mathbb E[\tilde A_i\tilde A_i^\top]e_ie_i^\top\right)  \nonumber \\
&=\text{tr}\left(n^{-1}\sum_{i=1}^n \mathbb E[\tilde A_i\tilde A_i^\top]\mathbb E[e_ie_i^\top]\right) + o_p(1)
   =n^{-1}\sum_{i=1}^n \mathbb E[e_i^\top\tilde A_i\tilde A_i^\top e_i]+o_p(1) = \sigma_{e,1}^2 + o_p(1), \label{eq: sigma hat estimation}
\end{align}
where the first equality follows from Assumption \ref{A4}(i), the second and the third from properties of the trace operator $\text{tr}(\cdot)$, the fourth from Assumptions \ref{A1}(i,iii), \ref{A4}(ii), and \ref{A5new}(i), the fifth from Assumption \ref{A4}(i) and properties of the trace operator $\text{tr}(\cdot)$, and the sixth from the definition of $\sigma_{e,1}^2$.

Next, let $(Z_1,Z_2)$ be a pair of independent standard normal random variables that is independent of everything else. Then for $\delta = \min(\delta_1,\delta_3)$,
\begin{align}
&\sup_{t\in\mathbb R}\left| \mathbb P\left( \hat\sigma_{e,1}^{-1}n^{-1/2}\sum_{i=1}^n\tilde A_i^{\top}e_i \leq t\mid \mathcal F_e \right) - \mathbb \mathbb P(Z_1\leq t)\right|  \nonumber \\
&\qquad\qquad \leq \frac{n^{-1-\delta/2}\sum_{i=1}^n\mathbb E[|\tilde A_i^\top e_i|^{2+\delta}\mid \mathcal F_e]}{\hat\sigma_{e,1}^{2+\delta}}
                          \leq \frac{K n^{-1-\delta/2}\sum_{i=1}^n \|e_i\|^{2+\delta}}{\hat\sigma_{e,1}^{2+\delta}} = o_p(1), \label{eq: conditional clt}
\end{align}
where the first inequality follows from a version of the Berry-Esseen theorem (see Section 35.1.9 in \cite{Dasgupta2008}), the second inequality from the Cauchy-Schwarz inequality and Assumptions \ref{A1}(iii) and \ref{A4}(ii), and the last bound from \eqref{eq: sigma hat estimation} and Assumptions \ref{A4}(iii) and \ref{A5new}(ii). Therefore, for any $t\in\mathbb R$,
\begin{align}
& \mathbb P(\sigma_{d\varepsilon}^{-1}(M_1 + M_2) \leq t) \nonumber\\
&\qquad  = \mathbb E[\mathbb P(\hat\sigma_{e,1}^{-1}M_1\leq \hat\sigma_{e,1}^{-1}(\sigma_{d\varepsilon}t - M_2)\mid\mathcal F_e)]
   = \mathbb E[\mathbb P(Z_1 \leq \hat\sigma_{e,1}^{-1}(\sigma_{d\varepsilon}t - M_2)\mid\mathcal F_e)] + o(1) \nonumber\\
&\qquad = \mathbb P(\hat\sigma_{e,1} Z_1 + M_2 \leq \sigma_{d\varepsilon}t) +o(1)
   = \mathbb P(\sigma_{e,1} Z_1 + M_2 \leq \sigma_{d\varepsilon}t) +o(1) \nonumber \\
&\qquad = \mathbb E[\mathbb P( \sigma_{e,2}^{-1}M_2 \leq \sigma_{e,2}^{-1}(\sigma_{d\varepsilon} t - \sigma_{e,1}Z_1) \mid Z_1 )] + o(1)
   = \mathbb E[\mathbb P( Z_2 \leq \sigma_{e,2}^{-1}(\sigma_{d\varepsilon} t - \sigma_{e,1}Z_1) \mid Z_1 )] + o(1) \nonumber \\
&\qquad = \mathbb P(\sigma_{d\varepsilon}^{-1}(\sigma_{e,1}Z_1 + \sigma_{e,2} Z_2)\leq t) + o(1) = \mathbb P(Z_1 \leq t) + o(1), \label{eq: proof of thm 2 end}
\end{align}
where the first equality follows from the law of iterated expectations (LIE), the second from \eqref{eq: conditional clt} by noting that the difference of two probabilities is always a number between zero and one to conclude that $o_p(1)$ in \eqref{eq: conditional clt} satisfies $\mathbb E[o_p(1)] = o(1)$, the third from the LIE, the fourth from \eqref{eq: sigma hat estimation} and Assumption \ref{A4}(iii), the fifth from the LIE, the sixth from Assumption \ref{A5new}(iii), the seventh from the LIE, and the eighth from noting that $\sigma_{e,1}Z_1 + \sigma_{e,2}Z_2$ is a normal random variable with mean zero and variance $\sigma_{d\varepsilon}^2 = \sigma_{e,1}^2 + \sigma_{e,2}^2$. This gives $\sigma_{d\varepsilon}^{-1}n^{-1/2}\sum_{i=1}^nd_i^*\varepsilon_i\to_d N(0,1)$ and completes the proof of the theorem. \hfill$Q.E.D.$

\bigskip

\noindent \textsc{Proof of Theorem \ref{L2}}. For this proof, it will be convenient to denote
$\mathbf{W}\equiv((w_{i,j}^{\top})_{i\leq n_{j}})_{j\leq n_{g}}$,
$\mathbf{Y}\equiv((y_{i,j})_{i\leq n_{j}})_{j\leq n_{g}}$, and $\bm{\epsilon}\equiv ((\varepsilon_{i,j})_{i\leq n_{j}})_{j\leq n_{g}}$. In addition, denote $\mathbf{D}\equiv
(d_{j}\mathbf{1}_{n_{j}\times1})_{j\leq n_{g}}$ and $\tilde{\mathbf{D}%
}\equiv \mathbf{D}-(n^{-1}\mathbf{D}^{\top}\mathbf{1}_{n\times1})$. Moreover, denote $\overline{d}_{n}\equiv n^{-1}\sum_{j=1}^{n_{g}}n_{j}d_{j}$.

As in the proof of Theorem \ref{L1}, given that the matrix $\mathbf W$ includes a non-zero constant column by Assumption \ref{A2}(iii), it follows that $\mathbf{1}_{n\times1}^{\top}\mathbf{M}_{W}=0$. Therefore, by the Frisch-Waugh-Lovell theorem,
\begin{equation}
\hat{\beta}-\beta=(\mathbf{D}^{\top}\mathbf{M}_{W}\mathbf{D})^{-1}%
(\mathbf{D}^{\top}\mathbf{M}_{W}%
\bm{\epsilon}%
)=(\mathbf{\tilde{D}}^{\top}\mathbf{M}_{W}\mathbf{\tilde{D}})^{-1}%
(\mathbf{\tilde{D}}^{\top}\mathbf{M}_{W}%
\bm{\epsilon}%
). \label{eq: thm 3 FWL theorem}%
\end{equation}
We first consider the denominator $\mathbf{\tilde{D}}^{\top}\mathbf{M}_{W}\mathbf{\tilde{D}}$. Since $n = \sum_{j\leq
n_{g}}n_{j}$, under Assumptions \ref{A1'}(i,iii) we have
\begin{equation}
\mathbb{E}\left[  |\overline{d}_{n}-\mu_d|^{2}\right]  =n^{-2}%
\sum_{j=1}^{n_{g}}n_{j}^{2}\mathbb{E}\left[  (d_{j}^{\ast})^{2}\right]  \leq
Kn^{-2}\sum_{j=1}^{n_{g}}n_{j}^{2} = K\kappa_n^2/n^2. \label{P_L2_1}%
\end{equation}
We therefore obtain from Markov's inequality that
\begin{equation}\label{eq: thm3 markov mean}
\overline{d}_{n}-\mu_d = O_p(\kappa_n/n),
\end{equation}
and so
\begin{equation}\label{P_L2_2}
\overline d_n^2 - \mu_d^2 = (\overline d_n - \mu_d)(\overline d_n + \mu_d) = O_p(\kappa_n/n).
\end{equation}
by Assumption \ref{A1'}(iii). In addition, again under Assumptions \ref{A1'}(i,iii) we have
$$
\mathbb E\left[\left|n^{-1}\sum_{j=1}^{n_g}n_j(d_j^2 - \mathbb E[d_j^2])\right|^2\right]
= n^{-2}\sum_{j=1}^{n_g}n_j^2\mathbb E\left[| d_j^2 - \mathbb E[d_j^2] |^2\right] \leq Kn^{-2}\sum_{j=1}^{n_g}n_j^2 =K\kappa_n^2 / n^2,
$$
and so
\begin{equation}\label{P_L2_3}
n^{-1}\sum_{j=1}^{n_g} n_j(d_j^2-\mathbb E[d_j^2]) = O_p(\kappa_n/n)
\end{equation}
by Markov's inequality.
Combining results (\ref{P_L2_2}) and (\ref{P_L2_3}) then yields%
\begin{equation}
n^{-1}\mathbf{\tilde{D}}^{\top}\mathbf{\tilde{D}}=n^{-1}\sum_{j=1}^{n_{g}%
}n_{j}(d_{j}-\overline{d}_{n})^{2}=n^{-1}\sum_{j=1}^{n_{g}}n_{j}d_{j}%
^{2}-(\overline{d}_{n})^{2}=\sigma_{d}^{2}+O_{p}(\kappa_n/n). \label{P_L2_4}%
\end{equation}
Further, since $d_{j}$ does not depend on $i$, we have
\begin{align}
n^{-1}\mathbf{\tilde{D}}^{\top}\mathbf{W}  &  =n^{-1}\sum_{j=1}^{n_{g}}%
d_{j}^{\ast}\sum_{i=1}^{n_{j}}w_{i,j}-\left(  n^{-1}\sum_{j=1}^{n_{g}}%
n_{j}d_{j}^{\ast}\right)  \left(  n^{-1}\sum_{j=1}^{n_{g}}\sum_{i=1}^{n_{j}%
}w_{i,j}\right) \nonumber \\
&  =n^{-1}\sum_{j=1}^{n_{g}}d_{j}^{\ast}\sum_{i=1}^{n_{j}}w_{i,j}+O_{p}%
(\kappa_n/n), \label{P_L2_5}%
\end{align}
where the second equality is by Assumption \ref{A2}(ii) and
(\ref{eq: thm3 markov mean}). Moreover, by Assumptions \ref{A2}(ii) and \ref{A1'}(i, ii, iii),%
\begin{align*}
\mathbb{E}\left[  \left \Vert n^{-1}\sum_{j=1}^{n_{g}}d_{j}^{\ast}\sum
_{i=1}^{n_{j}}w_{i,j}\right \Vert ^{2}\right]   &  =\sigma_{d}^{2}n^{-2}%
\sum_{j=1}^{n_{g}}\mathbb{E}\left[  \left \Vert \sum_{i=1}^{n_{j}}%
w_{i,j}\right \Vert ^{2}\right] \leq Kn^{-2}\sum_{j=1}^{n_{g}%
}n_{j}^{2} = K\kappa_n^2/n^2.%
\end{align*}
Therefore, by Markov's inequality we obtain
\[
n^{-1}\sum_{j=1}^{n_{g}}d_{j}^{\ast}\sum_{i=1}^{n_{j}}w_{i,j}=O_{p}(\kappa_n/n),
\]
which together with (\ref{P_L2_5}) further shows
that%
\begin{equation}
n^{-1}\mathbf{\tilde{D}}^{\top}\mathbf{W}=O_{p}(\kappa_n/n).
\label{P_L2_6}%
\end{equation}
Collecting the results (\ref{P_L2_4}), and (\ref{P_L2_6}) and using Assumptions \ref{A2}(i) and \ref{A3'}(iv),
we get%
\begin{equation}
n^{-1}\mathbf{\tilde{D}}^{\top}\mathbf{M}_{W}\mathbf{\tilde{D}}=n^{-1}%
\mathbf{\tilde{D}}^{\top}\mathbf{\tilde{D}}-n^{-1}\mathbf{\tilde{D}}^{\top
}\mathbf{W}(n^{-1}\mathbf{W}^{\top}\mathbf{W})^{-1}n^{-1}\mathbf{W}^{\top
}\mathbf{\tilde{D}}=\sigma_{d}^{2}+o_{p}(1). \label{P_L2_7}%
\end{equation}
Next, we consider the numerator $\mathbf{\tilde{D}}^{\top}\mathbf{M}_{W}%
\bm{\epsilon}$ in \eqref{eq: thm 3 FWL theorem}.
By\ (\ref{eq: thm3 markov mean}) and Assumptions \ref{A3'}(i) and \ref{A2}(iii),%
\[
\frac{\overline{d}_{n}-\mu_d}{S_{\varepsilon}}n^{-1/2}%
\sum_{j=1}^{n_{g}}\sum_{i=1}^{n_{j}}\varepsilon_{i,j}=\frac{O_{p}(\kappa
_{n}/n)}{S_{\varepsilon}}n^{-1/2}\sum_{j=1}^{n_{g}}\sum_{i=1}%
^{n_{j}}\varepsilon_{i,j}=o_{p}(1).
\]
Therefore,%
\begin{align}
\frac{n^{-1/2}\mathbf{\tilde{D}}^{\top}%
\bm{\epsilon}}{S_{\varepsilon}}  &  =\frac{n^{-1/2}\sum_{j=1}^{n_{g}}d_{j}^*\sum_{i=1}^{n_{j}}\varepsilon_{i,j}}{S_{\varepsilon}%
}-\frac{(\overline{d}_{n}-\mu_d)}{S_{\varepsilon}}%
n^{-1/2}\sum_{j=1}^{n_{g}}\sum_{i=1}^{n_{j}}\varepsilon_{i,j}\nonumber \\
&  =\frac{n^{-1/2}\sum_{j=1}^{n_{g}}d_{j}^*\sum_{i=1}%
^{n_{j}}\varepsilon_{i,j}}{S_{\varepsilon}}+o_{p}(1). \label{P_L2_8}%
\end{align}
In addition,
$$
\frac{n^{-3/2}\kappa_n\mathbf{W}^{\top}\bm{\epsilon}}{S_{\varepsilon}} = o_p(1)
$$
by Assumption \ref{A3'}(i),
which together with \eqref{P_L2_6} and (\ref{P_L2_8}) and Assumption \ref{A2}(i) shows
\begin{align}
\frac{n^{-1/2}\mathbf{\tilde{D}}^{\top}\mathbf{M}_{W}%
\bm{\epsilon}%
}{S_{\varepsilon}}  &  =\frac{n^{-1/2}\mathbf{\tilde{D}}^{\top}%
\bm{\epsilon}%
}{S_{\varepsilon}}-\frac{n^{-1/2}\mathbf{\tilde{D}}^{\top}%
\mathbf{W}(\mathbf{W}^{\top}\mathbf{W})^{-1}\mathbf{W}^{\top}%
\bm{\epsilon}%
}{S_{\varepsilon}}\nonumber \\
&  =\frac{n^{-1/2}\sum_{j=1}^{n_{g}}d_{j}^*\sum_{i=1}%
^{n_{j}}\varepsilon_{i,j}}{S_{\varepsilon}}+o_{p}(1). \label{P_L2_9}%
\end{align}
Given that $S_{\varepsilon
}^{-1}n^{-1/2}\sum_{j=1}^{n_{g}}(d_{j}-\mu_d)\sum_{i=1}^{n_{j}%
}\varepsilon_{i,j}=O_{p}(1)$ by Assumptions \ref{A1'}(i,iii) and \ref{A3'}(ii) and Markov's inequality and that $\sigma_d^2\geq K^{-1}$ by Assumption \ref{A1'}(iv), we obtain from \eqref{eq: thm 3 FWL theorem},
(\ref{P_L2_7}), and (\ref{P_L2_9}) that
$$
\frac{\sqrt n(\hat\beta - \beta)}{S_{\varepsilon}} = \frac{n^{-1/2}\sum_{j=1}^{n_{g}}d_{j}^*\sum_{i=1}%
^{n_{j}}\varepsilon_{i,j}}{\sigma_d^2 S_{\varepsilon}} + o_p(1)
$$
and so
\begin{equation}\label{eq: linearization thm 3 proof}
\frac{\sqrt n(\hat\beta - \beta)}{S_{\varepsilon}/\sigma_d} = \frac{n^{-1/2}\sum_{j=1}^{n_{g}}d_{j}^*\sum_{i=1}
^{n_{j}}\varepsilon_{i,j}}{\sigma_d S_{\varepsilon}} + o_p(1)
\end{equation}
by Assumption \ref{A1'}(iii).

We next derive the asymptotic distribution of $n^{-1/2}\sum_{j=1}^{n_{g}}d_{j}^*\sum_{i=1}
^{n_{j}}\varepsilon_{i,j}/(\sigma_d S_{\varepsilon})$. Let $\mathcal{F}_{j,n}$ denote the filtration
generated by $(\bm{\epsilon}^{\top},((d_{m})_{m\leq j})^{\top})$. Then by Assumptions \ref{A1'}(i,v),
$$
\mathbb{E}\left[  \frac{n^{-1/2}d_{j}^{\ast}\sum_{i=1}^{n_j}\varepsilon_{i,j}}{\sigma_{d}%
S_{\varepsilon}}|\mathcal{F}_{j-1,n}\right]
=\frac{n^{-1/2}\sum_{i=1}^{n_j}\varepsilon_{i,j}}{\sigma_{d}S_{\varepsilon}}\mathbb{E}\left[  \left.
d_{j}^{\ast}\right \vert \mathcal{F}_{j-1,n}\right]
=\frac{n^{-1/2}\sum_{i=1}^{n_j}\varepsilon_{i,j}}{\sigma_{d} S_{\varepsilon}}\mathbb{E}\left[  d_{j}%
^{\ast}\right]  =0
$$
almost surely, which implies that $n^{-1/2}d_{j}^{\ast}\sum_{i=1}^{n_j}\varepsilon_{i,j}%
/(\sigma_{d}S_{\varepsilon})$ is a martingale difference\ array with
respect to $\mathcal{F}_{j,n}$.\ Next, observe that Assumptions \ref{A1'}(i,v)
and \ref{A3'}(ii) yield
\begin{equation}
\sum_{j=1}^{n_g}\mathbb{E}\left[  \left(  \frac{n^{-1/2}d_{j}^{\ast}\sum_{i=1}^{n_j}
\varepsilon_{i,j}}{\sigma_{d}S_{\varepsilon}}\right)  ^{2}|\mathcal{F}%
_{j-1,n}\right]
=S_{\varepsilon}^{-2}n^{-1}\sum_{j=1}^{n_g}\left(\sum_{i=1}^{n_j}\varepsilon
_{i,j}\right)^{2}\rightarrow_{p}1. \label{P_L1_9 thm 3}%
\end{equation}
Moreover for any $\eta>0$, Assumptions \ref{A1'}(iii,iv,v) and \ref{A3'}(iii) allow us to
conclude that
\begin{align*}
&  \sum_{j=1}^{n_g}\mathbb{E}\left[  \allowbreak \left.  \left(  \frac
{n^{-1/2}d_{j}^{\ast}\sum_{i=1}^{n_j}\varepsilon_{i,j}}{\sigma_{d}S_{\varepsilon}}\right)
^{2}1\left \{  \left \vert \frac{n^{-1/2}d_{j}^{\ast}\sum_{i=1}^{n_j}\varepsilon_{i,j}}{\sigma
_{d}S_{\varepsilon}}\right \vert >\eta \right \}  \right \vert \mathcal{F}%
_{j-1,n}\right] \\
&  \leq \frac{1}{\eta^{\delta_4}}\sum_{j=1}^{n_g}\mathbb{E}\left[  \left.
\left \vert \frac{n^{-1/2}d_{j}^{\ast}\sum_{i=1}^{n_j}\varepsilon_{i,j}}{\sigma_{d}%
S_{\varepsilon}}\right \vert^{2+\delta_4}\right \vert
\mathcal{F}_{j-1,n}\right] \\
&  =\frac{1  }{\eta^{\delta_4
}(\sigma_{d}S_{\varepsilon})^{2+\delta_4}n^{1+\delta_4/2}}\sum_{j=1}%
^{n_g}\mathbb{E}\left[  |d_{i}^{\ast}|^{2+\delta_4}\right]\left|\sum_{i=1}^{n_j}\varepsilon_{i,j}\right|^{2+\delta_4}\leq \frac{K}{\eta^{\delta_4}
S_{\varepsilon}^{2+\delta_4}n^{1+\delta_4/2}}\sum_{j=1}^{n_g}\left|\sum_{i=1}^{n_j}\varepsilon_{i}\right|^{2+\delta_4} = o_p(1).
\end{align*}
Combining this bound with (\ref{P_L1_9 thm 3}), we can invoke the martingale
central limit theorem (see, e.g., Corollary 3.1 in \cite{HallHeyde1980}) to
conclude that%
\begin{equation}
\frac{n^{-1/2}\sum_{j=1}^{n_g}d_{j}^{\ast}\sum_{i=1}^{n_j}\varepsilon_{i,j}}{\sigma_{d}%
S_{\varepsilon}}\rightarrow_{d}N(0,1). \label{P_L1_12}%
\end{equation}
The claim of the theorem follows from combining this result with \eqref{eq: linearization thm 3 proof}.
\hfill$Q.E.D.$

\bigskip

\noindent \textsc{Proof of Theorem \ref{thm: clustering plus hetero}}.
As in the proof of Theorem \ref{L2}, we have
\begin{equation}\label{eq: linearization thm 4 beginning}
\frac{\sqrt n(\hat\beta - \beta)}{S_{\varepsilon}/\sigma_d} = \frac{n^{-1/2}\sum_{j=1}^{n_{g}}d_{j}^*\sum_{i=1}
^{n_{j}}\varepsilon_{i,j}}{\sigma_d S_{\varepsilon}} + o_p(1);
\end{equation}
see Equation \eqref{eq: linearization thm 3 proof} there and note that the derivation of \eqref{eq: linearization thm 3 proof} did not rely on Assumption \ref{A1'}(v), which we are not imposing here. Since $S_{d\varepsilon}\geq S_{e,1}$ and $S_{e,1}\geq S_{\varepsilon}K^{-1}$ by Assumption \ref{A8}(iii) and $\sigma_d \leq K$ by Assumption \ref{A1'}(iii), \eqref{eq: linearization thm 4 beginning} implies that
$$
\frac{\sqrt n(\hat\beta - \beta)}{S_{d\varepsilon}/\sigma_d^2} = \frac{n^{-1/2}\sum_{j=1}^{n_{g}}d_{j}^*\sum_{i=1}
^{n_{j}}\varepsilon_{i,j}}{S_{d\varepsilon}} + o_p(1),
$$
which yields the equality in \eqref{eq: clustering plus hetero convergence}.

We next derive the convergence result in \eqref{eq: clustering plus hetero convergence}, i.e. we show that $S_{d\varepsilon}^{-1}n^{-1/2}\sum_{j=1}^{n_{g}}d_{j}^*\sum_{i=1}^{n_{j}}\varepsilon_{i,j}\to_d N(0,1)$. To do so, we write
$$
n^{-1/2}\sum_{j=1}^{n_g}d_j^*\sum_{i=1}^{n_j} \varepsilon_{i,j} = n^{-1/2}\sum_{j=1}^{n_g}A_j^{\top}\sum_{i=1}^{n_j} e_{i,j} = \sum_{j=1}^{n_g}(A_j - \mu_A)^{\top}\sum_{i=1}^{n_j} e_{i,j} + n^{-1/2}\sum_{j=1}^{n_g}\mu_A^{\top}\sum_{i=1}^{n_j} e_{i,j}
$$
and denote the first and the second terms on the right-hand side by $M_1$ and $M_2$, respectively. Also, we denote $\tilde A_j\equiv A_j - \mu_A$ for all $1\leq j\leq n_g$. In addition, denote $\hat\S_{e,1}^2\equiv n^{-1}\sum_{i=1}^n \mathbb E[(\tilde A_j^\top e_{i,j})^2\mid \mathcal F_e]$, where $\mathcal F_e$ is the filtration generated by $((e_{i,j}^{\top})_{i\leq n_j})_{j\leq n_g}$.

Then $\hat S_{e,1}^2 - S_{e,1}^2 = o_p(S_{e,1}^2)$ by the same argument as that to used to derive \eqref{eq: sigma hat estimation} in the proof of Theorem \ref{L3} with Assumption \ref{A9}(i) replacing Assumption \ref{A5new}. In addition, letting $Z$ be a standard normal random variable that is independent of everything else, we have
$$
\sup_{t\in\mathbb R}\left| \mathbb P\left( \hat\S_{e,1}^{-1}n^{-1/2}\sum_{j=1}^{n_g}\tilde A_j^{\top}\sum_{i=1}^{n_j}e_{i,j} \leq t\mid \mathcal F_e \right) - \mathbb \mathbb P(Z\leq t)\right| = o_p(1)
$$
by the same argument as that used to derive \eqref{eq: conditional clt} in the proof of Theorem \ref{L3} with $\delta = 2$ and Assumption \ref{A9}(ii) replacing Assumption \ref{A5new}(ii). Finally,
$$
\mathbb P(S_{d\varepsilon}^{-1}(M_1+M_2)\leq t) = \mathbb P(Z\leq t) + o(1)
$$
for all $t\in\mathbb R$ by the same argument as that used to derive \eqref{eq: proof of thm 2 end} in the proof of Theorem \ref{L3} with Assumption \ref{A9}(iii) replacing Assumption \ref{A5new}(iii). This gives $S_{d\varepsilon}^{-1}n^{-1/2}\sum_{j=1}^{n_{g}}d_{j}^*\sum_{i=1}^{n_{j}}\varepsilon_{i,j}\to_d N(0,1)$ and completes the proof of the theorem.
\hfill$Q.E.D.$

\bigskip

\section{Asymptotic equivalence of two variance formula\label{APP_2}}

\begin{lemma}
\label{A_L2} Consider the linear regression model in (\ref{simple-regression}%
). Suppose: (i) $(d_{i})_{i\leq n}$ are i.i.d.\ with zero mean, finite and
nonzero variance $\sigma_{d}^{2}$; (ii) $(\varepsilon_{i})_{i\leq n}$ is
covariance stationary with auto-covariance function $\Gamma_{\varepsilon
}(\cdot)$ satisfying $\Gamma_{\varepsilon}(0)>0$ and $\sum_{j=1}^{\infty
}\Gamma_{\varepsilon}(j)^{2}<\infty$. For $\mathbf{D}\equiv(d_{i})_{i\leq n}$,
we then have
\begin{equation}
\frac{\mathbf{D}^{\top}\Omega \mathbf{D}}{\Gamma_{\varepsilon}(0)\mathbf{D}%
^{\top}\mathbf{D}}\rightarrow1\text{ almost surely as }n\rightarrow \infty,
\end{equation}
where $\Omega$ is the covariance matrix of $(\varepsilon_i)_{i\leq n}$.
\end{lemma}

\noindent \textsc{Proof of Lemma \ref{A_L2}}. Note that
\begin{equation}
n^{-1}\mathbf{D}^{\top}\Omega \mathbf{D}=n^{-1}\sum_{i_{1}=1}^{n}\sum_{i_{2}%
=1}^{n}d_{i_{1}}d_{i_{2}}\Gamma_{\varepsilon}(i_{1}-i_{2}) =\Gamma
_{\varepsilon}(0)n^{-1}\mathbf{D}^{\top}\mathbf{D}+2n^{-1}\sum_{i=2}^{n}U_{i}
\label{P_A_L2_1}%
\end{equation}
where $U_{i}\equiv \sum_{i^{\prime}=1}^{i-1}x_{i}x_{i^{\prime}}\Gamma
_{\varepsilon}(i-i^{\prime})$. Let $\mathcal{\tilde{F}}_{i}$ denote the
natural filtration generated by $(d_{j})_{j\leq i}$. Then, under the
assumption that $(d_{i})_{i\leq n}$ is i.i.d., it follows that $\{U_{i}%
,\tilde{\mathcal{F}}_{i}\}$ is a martingale difference sequence with variance
$\mathbb{E}[U_{i}^{2}]=\sigma_{d}^{4}\sum_{j=1}^{i-1}\Gamma_{\varepsilon
}(j)^{2}$. Therefore we obtain that
\begin{equation}
\sum_{i=2}^{n}i^{-2}\mathbb{E}\left[  U_{i}^{2}\right]  =\sigma_{d}^{4}%
\sum_{i=2}^{n}i^{-2}\sum_{j=1}^{i-1}\Gamma_{\varepsilon}(j)^{2}=\sigma_{d}%
^{4}\sum_{j=1}^{n-1}\Gamma_{\varepsilon}(j)^{2}\sum_{m=j+1}^{n}m^{-2}\leq
K\sigma_{d}^{4}\sum_{j=1}^{\infty}\Gamma_{\varepsilon}(j)^{2}<\infty
\label{P_A_L2_2}%
\end{equation}
where the first inequality follows from $\sum_{m=1}^{\infty}m^{-2}<K$ and the
last inequality is due to $\sigma_{d}^{2}<\infty$ and $\sum_{j=1}^{\infty}
\Gamma_{\varepsilon}(j)^{2}<\infty$ by assumption. Hence, \eqref{P_A_L2_2}
establishes that $\sum_{i=2}^{\infty}i^{-2}\mathbb{E}\left[  U_{i}^{2}\right]
<\infty$. By the martingale strong law of large numbers (see, e.g., Theorem
3.76 in \cite{white2014asymptotic}) we can therefore deduce that $n^{-1}%
\sum_{i=2}^{n} U_{i}\rightarrow0$ almost surely as $n\rightarrow \infty$, which
together with (\ref{P_A_L2_1}) yields
\begin{equation}
n^{-1}\mathbf{D}^{\top}\Omega \mathbf{D}-\Gamma_{\varepsilon}(0)n^{-1}%
\mathbf{D}^{\top}\mathbf{D}\rightarrow0\text{ almost surely as }%
n\rightarrow \infty. \label{P_A_L2_3}%
\end{equation}
Moreover, by condition (i) of the lemma and Kolmogorov's strong law of large
numbers (see, e.g., Theorem 3.1 in \cite{white2014asymptotic}), $n^{-1}%
\mathbf{D}^{\top}\mathbf{D}\rightarrow \sigma_{d}^{2}$ almost surely as
$n\rightarrow \infty$. Since $\sigma_{d}^{2}>0$ and $\Gamma_{\varepsilon}%
(0)>0$, the claim of the lemma then follows from (\ref{P_A_L2_3}%
).\hfill$Q.E.D.$

\end{document}